\RequirePackage{fix-cm}
\documentclass[twocolumn,epjc3]{svjour3}

\RequirePackage{color,graphicx,url}

\journalname{Eur. Phys. J. A}

\usepackage{amsfonts,amsmath,amssymb,bm,xspace}

\graphicspath{{./figures/}}

\usepackage[colorlinks=true]{hyperref}  

\graphicspath{{figures/}} 
\usepackage{multirow}
\usepackage{subcaption}
\usepackage{academicons} 
\newcommand{\orcid}[1]{\href{https://orcid.org/#1}{\textcolor[HTML]{A6CE39}{\aiOrcid}}}
\usepackage{xcolor}

\usepackage{bm}
\usepackage{enumitem}

\usepackage{nccmath}
\usepackage{microtype}
\usepackage{xfrac}
\usepackage{soul}

\hypersetup{%
    pdfsubject=Paper,
    pdfkeywords={nuclear physics} {MBPT} {chiral EFT} {asymmetric nuclear matter}
    unicode=true,
    breaklinks=true,
     colorlinks   = true,
     linkcolor = blue,
     citecolor = blue,
     menucolor = blue,
     urlcolor = blue
}
\newcommand{\bfvec}[1]{\bm{#1}}

\begin{document}
\title{NJL-Chiral Soliton and the Nucleon Equation of State at supra-saturation density: Impact of Chiral Symmetry Restoration }

\author{Bikram Keshari Pradhan\thanksref{e1,addr1}\href{https://orcid.org/0000-0002-2526-1421}{\includegraphics[scale=0.3]{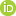}} 
\and
Guy Chanfray\thanksref{e2,addr1}\href{https://orcid.org/0000-0002-3593-9507}{\includegraphics[scale=0.3]{orcid.png}}
\and
Hubert Hansen\thanksref{e3,addr1}\href{https://orcid.org/0000-0001-8879-3612}{\includegraphics[scale=0.3]{orcid.png}}
\and
J\'er\^ome Margueron\thanksref{e4,addr2}\href{https://orcid.org/0000-0001-8743-3092}{\includegraphics[scale=0.3]{orcid.png}}
}
\institute{Institut de Physique des 2 infinis de Lyon, CNRS/IN2P3, Universit\'e de Lyon, Universit\'e Claude Bernard Lyon 1, F-69622 Villeurbanne Cedex, France \label{addr1}
\and
International Research Laboratory on Nuclear Physics and Astrophysics, Michigan State University and CNRS, East Lansing, MI 48824, USA\label{addr2}
}

\thankstext{e1}{b-k.pradhan@ip2i.in2p3.fr}
\thankstext{e2}{g.chanfray@ipnl.in2p3.fr}
\thankstext{e3}{hansen@ipnl.in2p3.fr}
\thankstext{e4}{jerome.margueron@cnrs.fr}

\date{Draft version:\today, Received: date / Accepted: date}

\maketitle

\begin{abstract}
It has been conjectured that, at sufficiently high baryon densities, the equation of state (EoS) of bulk nuclear matter can be identified with that of the nucleon core. In this work, we illustrate how the energy density and pressure distributions inside individual nucleons can be utilized to construct the EoS of supra-dense matter. In our framework,  nucleons arise as topological solitons stabilized by vector mesons, which are dynamically generated through the path integral bosonization of an underlying  Nambu–Jona-Lasinio (NJL)  model. The restoration of chiral symmetry is implemented dynamically via a self-consistent, density dependent scalar field, which modifies the (isovector) and (isoscalar) channels of the soliton. We analyze the resulting changes in soliton properties for different NJL parameter sets and demonstrate that the progressive restoration of chiral symmetry leads to a stiffening of the soliton-based EoS, making it compatible with existing neutron star EoSs. An EoS constructed from the solutions of the energy-density and pressure profiles at the center of the nucleon is also explored.

\end{abstract}

\section{Introduction}\label{sec:intro}
The phase transition of Quantum Chromodynamics (QCD) from nuclear matter into a quark-gluon plasma (QGP) is generally accompanied by chiral symmetry restoration and quark deconfinement~\cite{RISCHKE2004}. Theoretically, due to asymptotic freedom, perturbative QCD calculations are only applicable at short distances or high energy scales~\cite{Blaizot2003,Kajantie2003,Kurkela2010,Collins2011}.  At low densities and high temperatures, results from lattice QCD simulations indicate a smooth variation in thermodynamic properties, such as pressure ($p$), energy density ($\varepsilon$), and entropy with temperature. In contrast, phenomenological approaches like the Hadron Resonance Gas (HRG) model, which treats hadrons as non-interacting point particles (ideal gas) with the Hagedorn density of states~\cite{CABIBBO197567,Hagedorn:1967tlw,Ericson:2016sai,Broniowski2000}, predict a singular (blow-up) behavior in thermodynamic quantities near the Hagedorn temperature (chemical freeze-out). The singularity behavior is further tamed by excluded volume corrections that consider the finite hadron size, enabling a smooth crossover to quark matter~\cite{Andronic}, which suggests that the inter-hadron interactions play a crucial role. At high temperatures and densities, however, strong nucleon-nucleon interactions invalidate the HRG’s independent-particle picture, and as quark-gluon dynamics gradually dominate, the boundary between hadronic and partonic phases blurs. Hence, a theoretical unified description is demanding to account both hadronic and partonic degrees of freedom (d.o.f) in-order to describe  the thermodynamic properties of dense matter.

Though lattice QCD has significantly advanced our understanding of the equation of state (EoS) or the nature of the phase transition at low density and high temperature regime~\cite{Aoki2006,Cheng2008,Bernard2004,Bazavov2009,Ueda2014,Guenther2020}, it encounters the sign problem in the high density and low temperature regime~\cite{NAGATA2022}. A promising avenue to understand the  matter at high densities  lies in the study of astrophysical systems, in particular neutron stars (NSs). The densities realized in the inner cores of massive NSs exceed those achievable in terrestrial laboratories, making NSs unique natural laboratories for probing strongly interacting matter at extreme baryon densities (supra-dense matter). Although QCD cannot yet be solved directly at the relevant {density regimes}, the NS interior can be modeled through EoS based on microscopic many-body calculations or phenomenological approaches. These EoSs establish a connection between the microphysics of dense matter and global NS observables such as mass, radius, tidal deformability, and moment of inertia. Observational 
measurements
have significantly tightened constraints on the NS EoS. In particular, precise measurements of the massive $\sim 2M_{\odot}$ NS mass ~\cite{Demorest2010,Antoniadis:2013pzd,Cromartie:2019kug,Fonseca_2021} and the joint mass–radius measurements from the NICER mission~\cite{miller_m_c_2019_3473466,Riley_2019,Miller_2021,Riley_2021} have placed strong constraints on the range of viable models. Moreover, the ground-breaking detection of gravitational waves (GWs) from the binary NS merger event GW170817~\cite{AbbottPRX,Abbott_GWTC1,AbbottAJL848,AbbottPRL119,RICHABBOTT2021}, with its measurement of tidal deformability and NSs masses, has opened a new era of multi-messenger astronomy, providing complementary constraints on the dense matter EoS~\cite{AbbottPRL121}. Combined inputs from electromagnetic and GW observations have been widely employed to reduce EoS uncertainties and to explore the possible existence and nature of a hadron–quark phase transition in the interior of NSs~\cite{Most2018,Gloria2019,Somasundaram:2022,Roy2024,Christian2023a,Christian2023b,Pfaff2021,Somasundaram:2023a,HAN2023913}.
Moreover, it has been demonstrated that simultaneously incorporating the astrophysical constraints and/or the low-density constraints from chiral effective field theory (EFT) together with high-density limits from perturbative QCD provides information on the phase transition and novel constraints on the NS EoS~\cite{Shirke2022,Annala2019,Annala2022,Annala2023,Komoltsev2025,Gorda2023,Fan2024}, with the limitations expressed in Refs.~\cite{Somasundaram:2023b,Komoltsev2023}.

What if supra-dense matter already exists in finite-size nuclei? Nucleons, for instance, support densities that far exceed nuclear saturation density. In such a case, one should find a way to extract the thermodynamic properties of matter at these extreme densities. For instance several recent works have related the high-density environment of compact star cores to the mechanical properties of the nucleon, highlighting possible bridges between hadronic structure and NS matter \cite{Lorce2018,Fukushima2020}. The low-energy structure and dynamics of hadrons, including nucleons and pions, are governed by spontaneously broken chiral symmetry. Models based on chiral symmetry often view nucleons as a compact hard core that contains the three valence quarks and a surrounding quark–antiquark (pion) cloud. Furthermore, the idea of a compact core in the center of nucleons has also been supported by
deep-inelastic scattering data as well as from the evaluation of nucleon size by
high-energy nucleus-nucleus cross sections~\cite{HERA_2006,Xiong2019,Nijs2022}. 
The representation
of the nucleon as the compact hard core with a surrounding meson cloud also serves as the key idea for the cloudy MIT bag model ~\cite{Thomas1981,Miller2002} and the chiral soliton models ~\cite{ADKINS1984251,IGARASHI1985,REINHARDT1989825,REINHARDT1988577,Alkofer1994,Meissner1986,Meissner1987,Zuckert1994,Zuckert1995,Alkofer1995,Zuckert1997,Fukushima2020}. One central idea of this paper, following the work of Ref. \cite{Fukushima2020}, is to assume that at sufficiently high baryonic density, the thermodynamic pressure of bulk matter can be identified with the mechanical pressure in the hard core, in short, to identify the EoS of the nucleon core with the EoS of bulk matter. 

Since a rigorous derivation from QCD remains elusive, it is conjectured 
that baryons emerge as solitons. The Skyrme model, proposed initially by \cite{Skyrme1961}, describes baryons as topological solitons (Skyrmions) of the nonlinear pion field. While the model's kinetic term alone cannot support stable solutions due to Derrick's theorem~\cite{Derrick1964}, stability is achieved through the addition of the quartic term (Skyrme term). A significant refinement came with the inclusion of vector mesons, motivated by the Vector Meson Dominance (VMD) framework and the central role of the $\rho$ and $\omega$ mesons in $\pi\pi$, $\pi N$, and nucleon–nucleon interactions. In Ref.~\cite{ADKINS1984251}, it has been shown that coupling the $\omega$ meson alone can stabilize solitons without requiring the Skyrme term. At the same time, the inclusion of the $\rho$ meson has improved the description of the chiral soliton~\cite{IGARASHI1985}. The combined inclusion (axial) vector mesons $\omega$, $a$, and $\rho$ mesons, as discussed in Refs.~\cite{Meissner1986,Meissner1987}, provide a more comprehensive description of soliton dynamics and baryon properties. These vector meson extensions enhance the model's predictive power for nucleon structure, including electromagnetic form factors and baryon mass spectra, aligning closely with experimental observations. 

Numerous efforts have been devoted to derive effective hadron (nucleon and mesons) theories from the Nambu–Jona-Lasinio (NJL) model starting with the pioneering work of Ref.~\cite{EBERT1986188} and latter in Refs.~\cite{REINHARDT1990316,ISHII1995,BUCK1995168,Chanfray_PRC2011,Carrillo2014,CARRILLO2016,Noro2024,Chanfray2024EPJA,Chanfray2025}. These studies demonstrate that, at low energies, the structure of the effective meson Lagrangian is largely dominated by chiral symmetry. Bosonization of the NJL model yields an effective mesonic theory that reproduces the low-energy phenomenology of pseudoscalar (pion) and (axial-)vector mesons with reasonable accuracy. In addition, self-consistent soliton solutions of the bosonized NJL model have been extensively investigated to describe nucleons~\cite{REINHARDT1989825,REINHARDT1988577}. Various approaches have been developed to describe these solitons and to connect nucleon properties with experimental data, including the emergence of chiral solitons from the bosonized NJL model~\cite{REINHARDT1989825,REINHARDT1988577,Zuckert1994,Zuckert1995,Alkofer1995,Zuckert1997}, quark–diquark models~\cite{REINHARDT1990316,BUCK1995168,Bentz2001,Carrillo2014,CARRILLO2016}, quark–meson descriptions, and structure function methods~\cite{Weigel1997,Takyi2019}, as well as the appearance of $N$ and $N^*$ states in the parity-doublet model. Such soliton-based models
provide a transparent mechanism by which baryons can be understood as solitons of effective meson theories rooted in an underlying quark-level description. These NJL chiral soliton solutions can be compared with the Skyrme type model by a gradient expansion in the limit of infinitely heavy (axial-) vector mesons~\cite{Zuckert1995b}. Further extensions have also addressed the role of strangeness, providing insights into hyperons in the context of NJL-based chiral solitons~\cite{WEIGEL1994,Alkofer1995}.  

In-medium modifications of soliton properties, particularly in connection with the (fully or partial) restoration of chiral symmetry at finite density, have been investigated in several works, see Refs.~\cite{CHRISTOV1990,Alkofer1991,Amore2000}. In-medium effects are implemented through density dependent values of the pion decay constant as well as the in-medium pion and sigma masses in a quark-meson theory in Ref.~\cite{CHRISTOV1990}. It has ben suggested that the r.m.s charge radius of the soliton may increase by up to 17\% near nuclear saturation density $\rho_{\rm sat} \sim 0.16 \rm \  fm^{-3}$, and a phase transition from the chirally broken Goldstone phase to the chiral restored phase occurs near a baryon density of $\sim 3 \rho_{\rm sat}$~\cite{CHRISTOV1990}.

Another analysis suggests that the soliton solution vanishes near a critical quark chemical potential, indicating a phase transition where the bound soliton state dissolves to the constituents before the restored chiral symmetry~\cite{Alkofer1991}. Furthermore, within a soliton bag model theory in the Wigner-Seitz approximation to find the quark wavefunction and bound sigma fields, it was predicted that an equilibrium among the nuclear and quark phase could exist around $\sim 6\rho_{\rm sat}$ and a complete quark plasma phase could be expected at $\sim 12 \rho_{\rm sat}$~\cite{Birse1988}. Additionally, the soliton properties at finite temperature and density have also been examined, see Refs.~\cite{Alkofer1991,Abu2012,Mao2013,Li2024}. The soliton description of baryons with vector mesons has also been employed to understand the properties of nuclear matter, such as incompressibility and binding energy, both in the Wigner-Seitz cell and the multi-skrymion approach.  Since solitons are characterized by their r.m.s charge radii, this allows a self-consistent formalism  of incorporating finite-size effects into nuclear and NS matter, linking the evolution of baryon size directly to the underlying soliton description~\cite{Bentz2025}.

As discussed in Ref.~\cite{Fukushima2020}, the core and cloud picture provides a valuable framework to interpret baryonic structure at finite density. The outer, soft multi-pion cloud is directly linked to spontaneously broken chiral symmetry and the approximate Nambu–Goldstone nature of the pion. Its spatial extent is expected to grow with increasing density, consistent with the reduction of the in-medium pion decay constant. In contrast, the inner baryonic core is dominated by gluonic dynamics and is not directly tied to chiral symmetry, making it comparatively robust against moderate changes in density until the compact cores approach each other and begin to overlap. Well before this overlap sets in, the mesonic tails generate meson-exchange forces near saturation density ($\rho_{\rm sat}$). By further increasing the density, quark–antiquark pairs gradually delocalize, enhancing their mobility across multiple nucleons in a percolation-like process. This regime has been referred to as ``soft" deconfinement. At sufficiently high densities, however, when the baryonic hard cores themselves overlap, quark matter is realized unambiguously. This transition is termed ``hard deconfinement'' and may be identified with the deconfinement of valence quarks, at which point the properties of dense quark matter are governed primarily by quark degrees of freedom. It can be quantitatively assessed from the internal nucleon hard core substructure~\footnote{In Ref.~\cite{Fukushima2020}, the discussed nucleon EoS corresponds to the de-confined quark matter (valid near the hard core overlap $\sim 8 \rho_{\rm sat}$). However the EoS is represented as ``nucleon EoS" as it is derived from the core of the nucleon.}.

In the work presented in Ref.~\cite{Fukushima2020}, the EoS for dense quark matter in the hard-deconfined phase is derived from the local nucleon EoS associated with the soliton's hard core, constructed as a skyrmion of the pion field, evaluated under vacuum conditions. Although the restoration of chiral symmetry at high densities represents a natural physical expectation, its effects were not explicitly incorporated in the model. Instead, a qualitative argument was proposed to rescale the energy density and pressure using the in-medium pion decay constant as an approximation to account for chiral symmetry restoration. In contrast, the present study aims for a self-consistent approach to quantify the influence of chiral symmetry restoration on the density-dependent evolution of nucleon properties and the EoS. A central motivation of our framework is the description of the nucleon as a topological soliton emerging from the chiral dynamics of the NJL model, embedded within an in-medium effective background density-dependent scalar field that encapsulates the progressive restoration of chiral symmetry. {As described below, the parameters of this NJL model will be adjusted to reproduce the skyrmion solution under the vacuum condition used in Ref.~\cite{Fukushima2020}.}

To achieve the aim of this work, we leverage the path-integral bosonization approach of the underlying NJL model, as developed in Refs.~\cite{Chanfray2010,Chanfray2025}. In summary, the NJL model is bosonized to express quark-level interactions in terms of mesonic degrees of freedom, facilitating the description of hadronic properties. The effective Lagrangian of the NJL model, after bosonization, includes scalar, vector, and pseudoscalar meson fields. The formalism establishes direct connections between the NJL parameters and the vacuum and in-medium modifications to key meson properties, including the pion mass, pion decay constant, vector coupling constant, and vector meson masses, all of which are evaluated in terms of the density-dependent scalar field. This formulation enables us to have a systematic accounting of the connection among the NJL model parameters and to both vacuum and in-medium nucleon properties, such as the soliton's mass, radii, and EoS, while facilitating the exploration of scalar-field-induced or the restored chiral symmetry effects on mesons and solitons in dense nuclear matter.

This paper is organized as follows. In section~\ref{sec:Method_NJL}, the theoretical framework is provided, including the bosonized NJL model and the construction of the chiral soliton ansatz with vector mesons. In this section, details about the numerical implementation and its reliability by benchmarking it against existing results are presented as well. We discuss the sensitivity of the vacuum solution properties to the NJL model parameters and the effect of the chiral symmetry restoration on the nucleon structure and the resulting EoS in section~\ref{sec:results}. Finally, we summarise our findings and discuss the broader implications in section~\ref{sec:Conclusions}.

\section{Theoretical Framework}
\label{sec:Method_NJL}

In the present work, the nucleon is described within the NJL-soliton framework obtained through the bosonization of the NJL model, following the formulation developed in Refs.~\cite{Chanfray_PRC2011,Chanfray2025} and outlined in  section~\ref{sec:NJL_model}. The resulting effective mesonic Lagrangian provides the basis for the chiral soliton description of the nucleon, with the soliton stabilized by the  vector $\omega$ and vector isovector $\rho$ mesons. Our calculations are performed entirely within this NJL-soliton framework, without introducing an explicit confining interaction as considered in the original formulation of chiral confining model (CCM) in Refs.~\cite{Chanfray_PRC2011,Chanfray2025}. A description of the nucleon, including an explicit confining mechanism, will be addressed in future work.

\subsection{Bosonized NJL Model and Chiral Soliton Ansatz}\label{sec:NJL_model}

We start with defining the  NJL Lagrangian in Eq. \eqref{eq:NJL_Lagrangian}, within the light quark sector (especially $u$ and $d$ quarks within the exact isospin approximation), with the original aim of describing the low mass mesons such as the pion, the sigma, omega, and the rho mesons:
\begin{align}
{\cal L}_\mathrm{NJL}&= \overline{\psi}\left(i\,\gamma^{\mu}\partial_\mu\,-\,m_q\right)\,\psi\,\nonumber \\
&+\,\frac{G_1}{2}\,\left[\left(\overline{\psi}\psi\right)^2\,+\
\left(\overline{\psi}\,i\gamma_5\vec\tau\,\psi\right)^2\right]\nonumber \\
&-\,\frac{G_2}{2}\,\left[\left(\overline{\psi}\,\gamma^\mu\vec\tau\,\psi\right)^2\,+\,
\left(\overline{\psi}\,\gamma^\mu\gamma_5\vec\tau\,\psi\right)^2\,+\,\left(\overline{\psi}\,\gamma^\mu\,\psi\right)^2\right]~, \label{eq:NJL_Lagrangian} 
\end{align}
where $\psi$ are the quark fields. The model depends on three parameters, scalar coupling constants $G_1$, vector coupling constant $G_2$, and the current quark mass $m_q$. In addition to the three parameters from Eq.~\eqref{eq:NJL_Lagrangian}, a non-covariant cutoff parameter $\Lambda$ is introduced to avoid UV divergences in integrals, see below. In the absence of $G_2$, the parameters ($G_1, m_q,$ and $\Lambda$) are adjusted to reproduce the pion mass, the pion decay constant, and the quark condensate. The effect
of the meson parameters on $G_2$ is discussed later in this Section The choice of the parameters considered in this work is discussed in section~\ref{sec:results}.

The detailed methodology and the procedure of the bosonization of the NJL model used in this work have been developed in Refs.~\cite{Chanfray_PRC2011,Chanfray2025}. In summary, the bosonization of the NJL model in the current framework is done using path integral techniques to integrate the quark field, and after
a chiral rotation of the quark field, it can be equivalently written in a semi-bosonized form involving a pion field, embedded in the unitary operator $U=\xi^2$, a scalar field (${\cal S}$), a vector field ($\Omega_{\mu}$), vector iso-vector ($\vec{V}_{\mu}$), and an axial-vector field ($\vec{A}_{\mu}$)~\cite{Chanfray_PRC2011,Chanfray2025}.  The important virtue of the method lies in the fact that all the parameters entering the bosonized Lagrangian functionally depend on the value of the scalar field $\mathcal{S}$ whose vacuum expectation value coincides with the constituent quark mass $M_0\sim 350$~MeV. The amount of partial chiral symmetry restoration is associated with the decrease of the field $\mathcal{S}$. The value of $\mathcal{S}$, or its deviation $s$ from the vacuum value given by $s=(F_\pi/M_0)(\mathcal{S}-M_0)$, in the dense medium is obtained from the equation of motion for the scalar field as explained in section~\ref{sec:soliton_and_density}. It follows that all the quantities related to the nucleon properties should be seen as a functional of the nuclear scalar field $s$ induced by the presence of the other nucleons. In uniform matter, the value of the scalar field $\cal S$ is considered as uniform and equal to the value it takes for the density of nuclear matter. The fluctuation of the scalar field in and out of the nucleon is therefore disregarded by this prescription, which is in the spirit of the Born-Oppenheimer approximation in finite-size nuclei~\cite{GuichonNuPhA,GuichonPrPNP,Chanfray2024EPJA}. The value of the scalar field for a given nucleon is taken at the nucleon center of mass (CM) position $\bfvec{R}_N$ of the nucleon for the nuclear vector density $\rho(\bfvec{R}_N)$. For finite-size nuclei, $\mathcal{S}(\bfvec{R}_N)=(M_0/F_\pi)(F_\pi + s(\bfvec{R}_N))$.

In the current work, we describe the soliton with the pion, the vector, and the iso-vector mesons. Hence, the effective Lagrangian takes the form (after omitting axial meson contribution and the explicit scalar meson terms of the meson Lagrangian of Ref.~\cite{Chanfray_PRC2011}):
\begin{align}
{\cal L}_\mathrm{mes}&= \,\frac{m_q\,{\cal S}}{4\,G_1}\,\mathrm{Tr}_f(U\, +\, U^\dagger\,-\,2)\,\nonumber\\
&+\,
\tilde I({\cal S})\,{\cal S}^2\,\mathrm{Tr}_f\left({\cal A}^\mu_c\cdot{\cal A}^c_\mu\right)\nonumber \\
&+\,\frac{1}{4\,G_2}\,\mathrm{Tr}_f\left({V}^\mu +{\cal V}^\mu_c \right)^2\ \nonumber\\
& -\,\frac{1}{6}\,2 N_c N_f\,I_{2V}({\cal S})\,\left(\Omega^{\mu\nu}\Omega_{\mu\nu}\,+\,\vec{V}^{\mu\nu}\cdot \vec{V}_{\mu\nu}\,\right)\, ,   \label{eq:L_meson}
\end{align}
where $N_c$ is the number of colors and $N_f$ is the number of flavors. Furthermore in Eq.~\eqref{eq:L_meson},
\begin{eqnarray}
{\cal V}^\mu_c &=& \frac{i}{2}\left(\xi\partial^\mu\xi^\dagger\,+\,\xi^\dagger\partial^\mu\xi\right),\nonumber \\
{\cal A}^\mu_c&=&\frac{i}{2}\left(\xi\partial^\mu\xi^\dagger\,-\,\xi^\dagger\partial^\mu\xi\right) \nonumber \, .
\end{eqnarray}
The canonical vector mesons $\omega_{\mu}$ and vector isovector meson $\vec{\rho}_{\mu}$ are related by,
\begin{equation}
V^{\mu} =  {\Omega^\mu+\tau_j \vec{V}^{\mu}_j} \, ,
\end{equation}
\begin{equation}\
\Omega_\mu=g_v \omega_\mu ,\ \vec{V}_\mu=g_v \rho_{\mu} ~.
\end{equation}
The iso-vector coupling  $g_v=g_v({\cal S})$ is defined as:
\begin{equation}\label{eq:g_v}
 g_v^2({\cal S})=\frac{3/2}{2N_c N_f I_{2V}({\cal S})}~.   
\end{equation}
Given $\ E_{p}=\sqrt{{\cal S}^2+\bf p^2} $, the relevant integrals are defined as, 
\begin{align}
I_{2V}({\cal S})=&I_2({\cal S})+\frac{{\cal S}^2}{2}J_3({\cal S})-\frac{{\cal S}^4}{12}  J_4({\cal S})~, \\
I_2({S})=&\int_0^\Lambda \frac{d{\bf p}}{(2\pi)^3}\,\frac{1}{4\,E^3_p({S})},\ \nonumber \\
J_3({S})=&\int_0^\Lambda\frac{d{\bf p}}{(2\pi)^3}\,\frac{3}{8\,E^5_p({S})}, \nonumber\\
J_4({S})=&\int_0^\Lambda\frac{d{\bf p}}{(2\pi)^3}\,\frac{15}{16\,E^7_p({S})}~. \nonumber
\end{align}   
In Eq.~\eqref{eq:NJL_Lagrangian}, 
$$\tilde I({\cal S})\equiv 2 N_c N_f\,\tilde I_{2}({\cal S})=\frac{I({\cal S})}{1\,+\,4\,G_2\,{\cal S}^2\,I({\cal S})},$$
and,
$$I({\cal S})=2 N_c N_f\, I_{2}({\cal S})~.$$
Within the adopted bosonization framework with the path integral approach of the NJL model, {in-medium modified} pion decay constant parameter  $F_{\pi, \rm NJL} ({\cal S})$, mass of the pion $M_{\pi, \rm NJL} ({\cal S})$, mass of the scalar meson $M_{\sigma}({\cal S})$ and mass of the vector meson $M_v ({\cal S})$ an be expressed as follows:
\begin{equation}\label{eq:Mpi}
M_{\pi,\rm NJL}^2({\cal S}) =
\frac{m_q {\cal S}}{G_1 F^2_{\pi,\rm NJL}({\cal S})}~,
\end{equation}
\begin{align}
F^2_{\pi, \rm NJL}({\cal S}) &=
\frac{f_{\pi,G_2=0}^2}{1 + 4 G_2 f_{\pi,G_2=0}^2}~, \label{eq:Fpi} \\
f_{\pi, G_2=0}^2 &=
2 N_c N_f {\cal S}^2 I_2({\cal S})~,\nonumber
\end{align}
\begin{align}
M^2_{\sigma, \rm NJL}({\cal S}) &=
\left(1 + 4 G_2 f_{\pi,G_2=0}^2\right)
\left(4{\cal S}^2 + m_{\pi, \rm NJL}^2\right)~, \label{eq:Msigma} \\
m_{\pi, \rm NJL}^2 &=
\frac{{\cal S} m_q}{G_1 f_{\pi,G_2=0}^2}~,\nonumber
\end{align}
\begin{align}
M_{v,\rm NJL}^2({\cal S}) &=
\frac{g_v^2({\cal S})}{G_2}
= 4 a({\cal S}) g_v^2({\cal S}) F_{\pi,\rm NJL}^2({\cal S})~,
\label{eq:Mv} \\
a({\cal S}) F_{\pi,\rm NJL}^2({\cal S}) &=
\frac{1}{4G_2}~. \label{eq:a_of_s}
\end{align}
Using the definitions of the meson parameters in terms of the scalar field ${\cal S}$ and the addition of the Wess-Zumino term ($\cal L_{\rm WZ}$) for the  $\omega$-meson coupling to the conserved baryon current $B^{\mu}$, the Lagrangian density of the soliton can be expressed as,
\begin{align}
{\cal L}_{\rm slt}&= \,\frac{1}{4}F^2_{\pi}({\cal S})M^2_{\pi}({\cal S})\,\ \mathrm{Tr}_f(U\, +\, U^\dagger\,-\,2)\,\nonumber \\
&+\, \frac{F_{\pi}^2({\cal S})}{4}
\mathrm{Tr}_f\left(\partial_{\mu} U\partial^{\mu} U^{\dagger}\right)
\, \nonumber \\
&+\,a({\cal S})  F_{\pi}^2({\cal S})\,\mathrm{Tr}_f\left({V}^\mu +{\cal V}^\mu_c \right)^2\nonumber\\
& -\,\frac{1}{4}\,\left(\omega^{\mu\nu}\omega_{\mu\nu}\,+\,\vec{\rho}^{\mu\nu}\cdot \vec{\rho}_{\mu\nu}\right) +\cal L_{\rm WZ} ~. \label{eq:L_meson2}
\end{align}
In ~Eq.~\eqref{eq:L_meson2}, the term $\cal L_{\rm WZ}$ is defined as,
\begin{eqnarray}
  {\cal L}_{\rm WZ}=N_c g_v(S) \omega_\mu B^{\mu} \, ,
\end{eqnarray}
where $B^{\mu}$ is the baryon charge current given as,
\begin{equation}
B^{\mu}=\frac{1}{24\pi^2}\epsilon^{\mu \nu \alpha \beta} \ \mathrm{Tr} \left[ U^{\dagger}\partial_\nu U\ . \ U^{\dagger}\partial_\alpha U \ . \ U^{\dagger}\partial_\beta U\right]~.
\end{equation}

The solitons considered in this work are constructed within a bosonized NJL framework using a derivative expansion, whose domain of validity is well understood~\cite{Zuckert1995b}. Although this approximation is not systematically controlled when the soliton size approaches the scale of integrated-out physics, it provides a standard and widely used description of the leading mesonic dynamics responsible for soliton formation. While more refined treatments have been developed to address this limitation (see for instance ~\cite{Diakonov1997a,Diakonov1997b,Goeke2001}), their implementation lies beyond the scope of the present work. However, our aim here is not to refine or study the detailed chiral soliton model itself, but to illustrate a physical mechanism within a simple and tractable framework (a framework that has been adopted in ~\cite{Fukushima2020}) that retains the essential chiral symmetry properties.

In the NJL-soliton framework, we construct baryons as Skyrmions within a chiral model incorporating $\pi$, $\rho$, and $\omega$ meson fields describing the baryon structure. For simplicity, we focus on static solutions, neglecting the (iso-)spin quantization of the soliton. Unlike the standard Skyrme model, which relies on a higher-order Skyrme term for stable soliton solutions, here we focus on the solitons stabilized by the vector $\omega$ and iso-vector $\rho$  mesons~\cite{ADKINS1984251,IGARASHI1985,Meissner1986}. Furthermore, to describe the soliton, we consider the hedgehog ansatz for the pion field, with corresponding configurations for the vector mesons, specifically the Wu-Yang-’t Hooft-Polyakov ansatz for the $\rho$ meson and a time-component-only form for the $\omega$ meson due to the absence of spatial source terms in static solutions:
\begin{align}
  U(\mathbf{r}) &= e^{i \boldsymbol{\tau} \cdot \hat{\mathbf{r}} F(r)}, \\
  \rho^{i,a}(\mathbf{r}) &= \epsilon^{ika} \hat{\mathbf{r}}^k \frac{-G(r)}{g r}, \\
  \omega^\mu(\mathbf{r}) &= \delta^{\mu 0} \omega(r),
\end{align}
where $\boldsymbol{\tau}$ denotes the Pauli matrices, $F(r)$ is the chiral angle, $\epsilon^{ika}$ is the Levi-Civita symbol with $i$ and $a$ representing spatial and isospin indices, respectively, and $G(r)$ and $\omega(r)$ are the radial profiles of the $\rho$ and $\omega$ mesons. Given the Lagrangian density in~Eq.~\eqref{eq:L_meson2}, and the particular choices of the meson profiles, the soliton energy density ($\varepsilon_{\rm{slt}}$)  can be obtained as outlined in Refs.~\cite{Meissner1986,Meissner1987,IGARASHI1985,Ma2013};
\begin{align}
\varepsilon_{\rm{slt}}(r) &= F_{\pi}^2({\cal S}) \left( \frac{1}{2} (F')^2 + \frac{\sin^2 F}{r^2}\right) \nonumber \\
&+  F_{\pi}^2({\cal S}) \left(\frac{a ({\cal S}) (G + 1 - \cos F)^2}{r^2} \right) \nonumber \\
&+ \frac{G'^2}{4 g_v^2({\cal S}) r^2} + \frac{G^2 (G + 2)^2}{8 g_v^2({\cal S}) r^4} \nonumber \\
& - \frac{1}{2} M_v^2({\cal S}) \omega^2 - \frac{1}{2} (\omega')^2 + \frac{3 g_v({\cal S})}{2 \pi^2 r^2} \omega F' \sin^2 F ~\nonumber \\
&+ M_{\pi}^2({\cal S}) F_{\pi}^2({\cal S})\  (1 - \cos F)~. \label{eq:ener}
\end{align}
 As pointed out in Ref.~\cite{Fukushima2025}, there is a pseudo-gauge ambiguity impacting the nucleon EOS. However, in this work, we follow the pragmatic prescription of ~\cite{Fukushima2020} by employing a mixture of the canonical and Belinfante energy momentum tensor (EMT), which removes the  singularity in the pressure (see Ref.~\cite{Fukushima2025} for details).  Within this framework and consideration, the soliton pressure ($p_{\rm slt}$) is calculated as in ~\cite{Fukushima2020}:
\begin{align}
p_{\rm{slt}}(r) &= -\frac{F_{\pi}^2({\cal S})}{3 r^2}\left[ \frac{1}{2} (rF')^2 + \sin^2 F \right] \nonumber \\
&-\frac{F_{\pi}^2({\cal S})}{3 r^2}\left[ a({\cal S}) (G + 1 - \cos F)^2 \right] \nonumber \\
&+ \frac{(G')^2}{12 g_v^2({\cal S}) r^2} + \frac{G^2 (G + 2)^2}{24 g_v^2({\cal S})r^4} \nonumber \\
 &+ \tfrac{1}{2} M_v^2({\cal S})\, \omega^2 - \tfrac{1}{6} (\omega')^2 \nonumber \\
&- M_{\pi}^2 ({\cal S}) \, F_{\pi}^2({\cal S}) \, (1 - \cos F)~.\label{eq:pressure}
\end{align}

In Eqs.~\eqref{eq:ener} and ~\eqref{eq:pressure}, $F_{\pi}({\cal S})$, $M_{\pi}({\cal S})$, $M_v({\cal S})$, $g_v({\cal S})$, and $a({\cal S})$ are related to the NJL model parameters and the scalar field $S$ as described in Eqs.~\eqref{eq:Fpi}, \eqref{eq:Mpi}, \eqref{eq:Mv}, \eqref{eq:g_v}, and \eqref{eq:a_of_s}, respectively. For clarity and simplicity, we have omitted the subscript $\rm NJL$ afterwards when representing the meson parameters, similar to those in Eqs.~\eqref{eq:ener} and ~\eqref{eq:pressure}. 

Note that ${\cal S}$ is the scalar field and in vacuum it coincides with the constituent quark mass $M_0$ as the solution of the Gap equation given in Eq.~\ref{eq:gap},
\begin{equation}\label{eq:gap}
M_0=m_q+4N_cN_f M_0 G_1 I_1(M_0) \, ,
\end{equation}
whereas in dense matter, the in-medium modification to the scalar field ${\cal S}$ in the context of nuclear physics is characterized by defining the  ``nuclear physics" scalar field `$s$' as:
\begin{equation}\label{eq:nuclear_s_to_S}
{\cal S}\equiv\frac{M_0}{f_{\pi}}(s+f_{\pi})\, .
\end{equation}

The scalar field ${\cal S}$ is introduced as a chiral-invariant effective field that parametrizes medium-induced modifications of the underlying chiral dynamics. It is worth mentioning  that the scalar field ($\cal S$) is not an order parameter of chiral symmetry breaking, unlike the quark condensate, whose density dependence arises from  both the scalar mean field and the pionic scalar density. Nevertheless, ${\cal S}$ remains a useful and physically meaningful indicator of the in-medium evolution of  chiral dynamics, as it is closely related to the constituent quark mass in NJL-based approaches. Hence, in the present framework, medium effects associated with partial chiral symmetry restoration enter the soliton description through the density dependence of $\cal S$. As the density increases and chiral symmetry is progressively restored  (corresponding to $ |s| \to f_{\pi}  $ or ${\cal S} \to 0 $ ), the in-medium pion and scalar masses approach each other ( see discussions in section~\ref{sec:soliton_and_density} and ~\ref{app:mesons_in_nuclear_matter}). This signals the gradual degeneracy of chiral partners and reflects the progressive restoration of chiral symmetry. This behavior is consistent with general expectations from chiral symmetry restoration and provides an interpretation of the role played by $\cal S$ in the model.

Throughout this article, $f_{\pi}\equiv F_{\pi, \rm NJL} ({\cal S}=M_0)$ represents the pion decay constant parameter in the vacuum if not mentioned elsewhere. The notation is chosen to be compatible with the general notation.

\subsection{Field Equations and Numerical Method}\label{sec:Numerical_method}

The solutions of the meson profiles can be obtained by minimizing the total energy ($E_{\rm slt}$) or the mass of the soliton $M_{\rm{slt}}=E_{\rm slt}$ given by:
\begin{equation}\label{eq:Mass}
M_{\rm{slt}}=E_{\rm slt}=4 \pi \int r^2 \varepsilon_{\rm{slt}}(r, \phi = \{F, G, \omega\}) \, dr~.
\end{equation}
The variational principle yields the equations of motion for the meson fields $\phi = \{F, G, \omega\}$:
\begin{align}
F^{\prime \prime} &= -\frac{2 F^{\prime}}{r}+ \frac{\sin{2F}}{r^2}[1-{a({\cal S})}]+\frac{2 {a({\cal S})}\sin F}{r^2} (G+1)\nonumber~ \\
 &+ {M_{\pi,\text{NJL}}^2({\cal S})} \sin F - \frac{6 g_v({\cal S})}{4 \pi^2r^2 {F^2_{\pi,\text{NJL}}({\cal S})}} \omega^{\prime} \sin^2 F ~, \label{eq:d2F_NJL}\\
G^{\prime \prime} &= {4g_v({\cal S})^2} {F_{\pi,\text{NJL}}^2({\cal S}) \ a({\cal S})} \ (G + 1 - \cos F)\nonumber\\
& + \frac{G (G + 2) (G + 1)}{ r^2} ~ ,\label{eq:d2G_NJL} \\
\omega^{\prime \prime} &= -\frac{2 \omega^{\prime}}{r}+ {M_v^2({\cal S})} \omega - \frac{6 g_v({\cal S})}{4 \pi^2 r^2} F^{\prime} \sin^2 F \label{eq:d2w_NJL}~ .
\end{align}

The baryon charge density $\rho_B(r)$ due to valence quarks, localized in the soliton core, is:
\begin{align}
\rho_B(r) &= -\frac{1}{2 \pi^2 r^2} F'(r) \sin^2 F(r), \label{eq:rhob}
\end{align}
with the quantized baryon charge $B = \int_0^\infty 4 \pi r^2 \rho_B(r) \, dr$. For $B = 1$ solutions and ensuring finite energy  of the soliton Eq.~\eqref{eq:ener}, the boundary conditions are:
\begin{align}
F(0) &= \pi, \quad F(\infty) = 0, \\
G(0) &= -2, \quad G(\infty) = 0, 
\end{align}
and,
\begin{align}
    \omega'(0) &= 0, \quad \omega(\infty) = 0,
\end{align}

The iso-scalar charge density is then defined as:
\begin{equation}
\rho_S(s) \propto -\omega(r)~.\label{eq:isocharge}
\end{equation}
The proportionality constant in Eq.~\eqref{eq:isocharge}, does not depend upon $r$ and is adjusted to normalize the iso-scalar density ($\int  \ \rho_S(r) \ d^3r=1$). To extract nucleon properties, one must solve the coupled second-order differential equations given in Eqs.~\eqref{eq:d2F_NJL},~\eqref{eq:d2G_NJL}, and ~\eqref{eq:d2w_NJL}.) subject to boundary conditions at the origin and at spatial infinity. The known initial values are$F(0) $,$G(0) $, and $\omega'(0) $, while the derivatives $F'(0) $, $G'(0) $, and the initial value $\omega(0) $ must be determined iteratively. A common strategy is to use the shooting method: integrating outward from $r = 0$  with trial values for the unknowns, adjusting them until the boundary conditions at large $r$ are satisfied.

In practice, this system is highly stiff and extremely sensitive to initial conditions. Even a small deviations in the trial parameters lead to divergent or oscillatory (“frog-like”) behavior, hindering the ability to reach physically meaningful solutions at $r \gtrsim 3\,\mathrm{fm} $. Although implicit methods offer better control over instabilities compared to explicit methods, such as the Runge–Kutta (RK4) method, they still fail to prevent exponential growth in the solutions. Moreover, even when a stable profile is achieved up to a finite matching point $r_\infty$ with suitable values of $F'(0)$, $G'(0)$, and $\omega(0)$, extending the integration often results in divergent behavior. While we can reproduce the soliton solutions presented in Ref.~\cite{Fukushima2020} using the shooting method with fine-tuned initial conditions, the system's extreme sensitivity renders the shooting method impractical for systematic studies.

The system of coupled, second-order differential equations for the profile functions \( F(r) \), \( G(r) \), and \( \omega(r) \) constitutes a mixture of initial value problem and boundary value problem (somewhat similar in structure to those encountered in quantum mechanics, such as the inverse-cusp potential). Due to the stiffness and sensitivity of the equations, conventional shooting methods prove inadequate. Instead, we employ a relaxation method from Ref.~\cite{numerical}, which is naturally suited for solving such problems.

\subsection{ Numerical Implementations}
\label{sec:Numerical_method_Testing}

Before going into the vast connection between nucleon properties of the NJL model parameters, we establish the reliability of our numerical framework for the NJL-soliton model,  validating our implementation by reproducing the results of Ref.~\cite{Fukushima2020}. The skyrmion framework, as detailed in Ref.~\cite{Fukushima2020}, provides soliton solutions in vacuum. In our NJL soliton framework, this is equivalently characterized by the scalar field ${\cal S} = M_0$. In this section, we adopt the below-mentioned {meson parameters} {to recover the skyrmion framework} from Ref.~\cite{Fukushima2020} to ensure consistency and direct comparison to their findings. Specifically, we use the following meson parameters to obtain the soliton solutions at vacuum following the methodology given in section~\ref{sec:NJL_model}:
\begin{align}
&g_v({\cal S}=M_0) \to \  g/2 =3~,\\
&a({\cal S}=M_0) \to\  2~,\\
&F_{\pi,\rm NJL}({\cal S}=M_0) \to\  f_{\pi}~,\\
&M^2_v({\cal S}=M_0) \to \ m^2_{\rho}=m^2_{\omega}={2g^2f_{\pi}^2}~,
\end{align}
and, 
\begin{align}
M_{\pi,\rm NJL}({\cal S}=M_0)&\to \  M_{\pi}~.
\end{align}
Furthermore, for the vector and iso-vector mesons, we use $m_{\omega} = m_{\rho} = 783$ MeV and $g = 6$. These substitutions align the NJL-soliton equations, see Eqs.~\eqref{eq:ener},~\eqref{eq:d2F_NJL}, and \eqref{eq:d2G_NJL}, with the expressions derived in Ref.~\cite{Fukushima2020,Meissner1987}, enabling a direct comparison.

In the relaxation method, we first employ a compactified coordinate transformation, $t = \tanh(r) \in [0,1]$ (where $r$ is expressed in fm), to handle the boundary condition at $r \to \infty$ and speed up the convergence. This transformation ensures numerical stability,  despite the rapid variation of the tanh function, as the mapping allows us to reach $r_{\infty} \sim 10$~fm for $1 - t \sim 10^{-9}$, which sufficiently captures the soliton tail, given that the characteristic nucleon radius is $\sim 1$~fm. A second-order central finite differences approximation is considered for the derivatives in the discretized coordinate system. The differential Eqs.~\eqref{eq:d2F_NJL},~\eqref{eq:d2G_NJL}, and~\eqref{eq:d2w_NJL} were reformulated as a nonlinear algebraic system, and a residual vector was constructed by subtracting the right-hand sides of the field equations Eqs.~\eqref{eq:d2F_NJL},~\eqref{eq:d2G_NJL}, and~\eqref{eq:d2w_NJL}  from the finite-difference approximations. The resulting nonlinear algebraic system is solved iteratively by minimizing the residual vector constructed from the discretized field equations. For initial guesses, we consider physically motivated profile functions inspired by prior literature. Exponentially decaying forms such as \( F(r) = \pi e^{-M_\pi r} \), \( G(r) = -2 e^{-M_v r} \), and \( \omega(r) = \omega_0 e^{-M_v r} \) yield reasonable convergence. However, we observe faster and more stable convergence with tanh-type trial functions: \( F(r) = \pi (1 - \tanh(M_\pi r)) \), \( G(r) = -2 (1 - \tanh(M_v r)) \), and \( \omega(r) = \omega_0 (1 - \tanh(m_\omega r)) \). These satisfy the boundary conditions and exhibit the correct asymptotic behavior implied by Eqs.~\eqref{eq:d2F_NJL},~\eqref{eq:d2w_NJL}.

\begin{figure}[htbp]
\centering
\includegraphics[width=\linewidth]{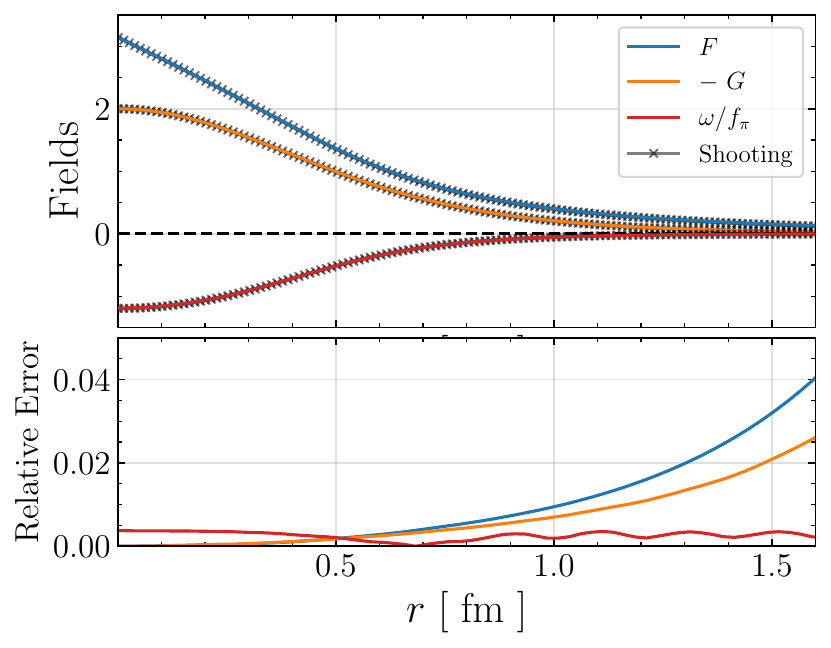}
\caption{The upper panel shows the field profiles obtained with the relaxation method (solid lines) compared with those from the shooting method (grey crosses), while the lower panel displays the relative error of the relaxation results with respect to the shooting method. The colour scheme for the different fields is kept consistent across both panels.}
\label{fig:Fields_fukushima}
\end{figure}

To compute the soliton’s physical properties in the radial domain with minimal errors, we used the converged solution in the compactified coordinate $t = \tanh(r)$ as an initial guess for the relaxation method in $r$.  Because we begin with a smooth, converged solution in $t$, convergence in $r$ is both stable and precise resulting as a polishing algorithm. A thorough numerical accuracy and error analysis is presented in ~\ref{app:error}. The meson field profiles, obtained using both the shooting and relaxation methods, are presented in Fig.~\ref{fig:Fields_fukushima}. As shown in~Fig.~\ref{fig:Fields_fukushima}, the field profiles obtained with the relaxation method agree with those obtained from the shooting method within $\leq 5\%$ over the relevant soliton radius range. The excellent agreement between these results confirms the robustness of our implementation.

\begin{figure}[htbp]
\centering
\includegraphics[width=\linewidth]{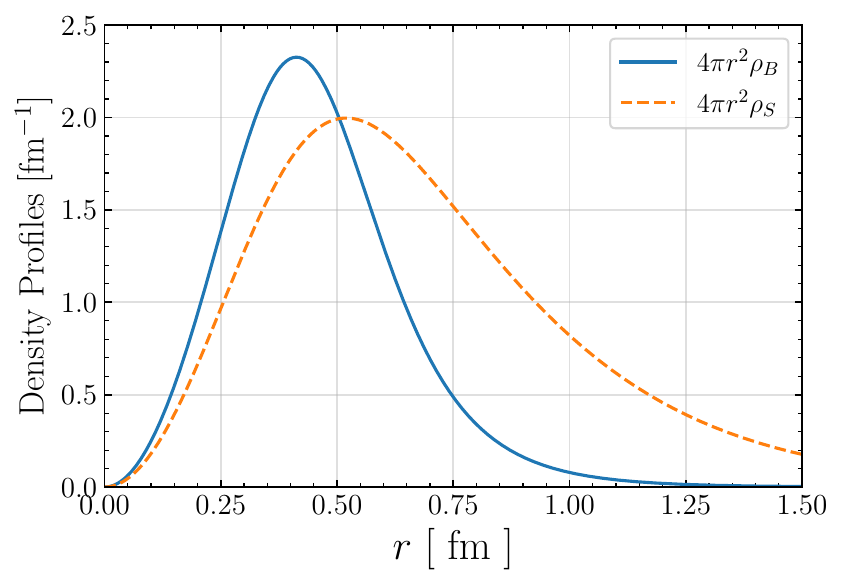}
\caption{Baryon ($\rho_B(r)$) and isoscalar ($\rho_S(r)$) charge densities, scaled by $4\pi r^2$, as functions of $r$. The radial axis is limited to $r_{\rm max}=1.5$~fm for visibility and comparison with Ref.~\cite{Fukushima2020}. For clarity and comparison with Ref.~\cite{Fukushima2020}, the isoscalar charge density is shown as $\rho_S(r)=-2gF_{\pi}^2\omega(r)$.}
\label{fig:density_vs_r_Fukushima}
\end{figure}

Using the parameters from Ref.~\cite{Fukushima2020}, we computed the normalized baryon and isoscalar charge density profiles, shown in Fig.~\ref{fig:density_vs_r_Fukushima}. The baryon density $\rho_B(r)$ reflects the valence quark contributions, defining the hard-core radius $r_B$, while the isoscalar density $\rho_S(r)$ captures the sea quark contributions, characterizing the isoscalar charge radius $r_S$.

The characteristic length scales were calculated as:
\begin{align}
\sqrt{\langle r^2_B \rangle} &= \left( \frac{\int_0^{r_{\infty}} dr\, r^4\, \rho_B(r)}{\int_0^{r_{\infty}}dr\, r^2\, \rho_B(r)} \right)^{1/2} \approx 0.499 \, \text{fm}, \\
\sqrt{\langle r^2_S \rangle} &= \left( \frac{\int_0^{r_{\infty}} dr\, r^4\, \rho_S(r)}{\int_0^{r_{\infty}} dr\, r^2\, \rho_S(r)} \right)^{1/2} \approx 0.796 \, \text{fm}.
\end{align}

The baryon charge radius, $\sqrt{\langle r^2_B \rangle} \approx 0.49$ fm, agrees well with Ref.~\cite{Fukushima2020}. However, a discrepancy arises in the isoscalar charge radius, where we obtain $\sqrt{\langle r^2_S \rangle} \approx 0.80$ fm, compared to $1.03$ fm reported in Ref.~\cite{Fukushima2020}. This difference, approximately 22\%, can be reconciled using the relation among $\langle r^2_B \rangle$ and $\langle r^2_S \rangle$ within the form factor argument as given in Refs.~\cite{Meissner1987,Fukushima2020,ADKINS1984251}:
\begin{equation}
\langle r^2_S \rangle = \langle r^2_B \rangle + \frac{6}{m_\omega^2}.
\end{equation}
with $m_\omega = 783$~MeV, we find  $\langle r^2_S \rangle \approx 0.63~\mathrm{fm}^2$, or $\sqrt{\langle r^2_S \rangle} \approx 0.8$~fm consistent with our result and with the earlier findings in Ref.~\cite{Meissner1987} (see section 4.4 therein) which is the foundation of the model used in Ref.~\cite{Fukushima2020}.

\begin{figure}[t]
\centering
\includegraphics[width=\linewidth]{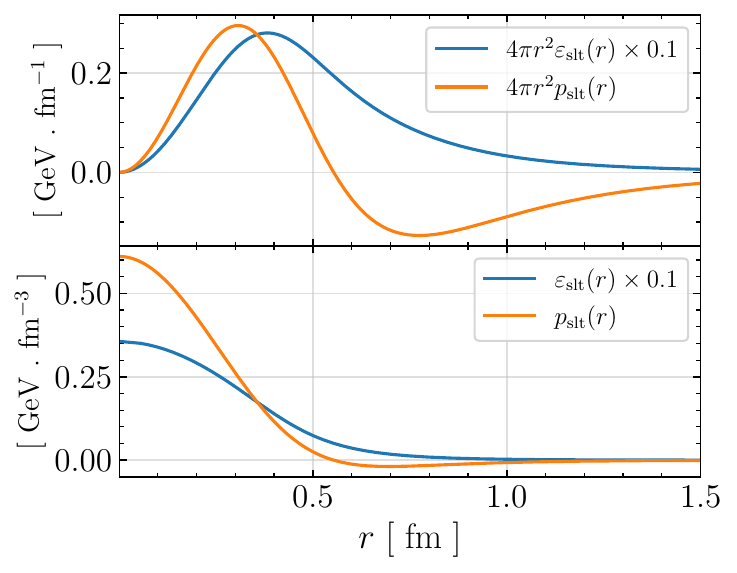}
\caption{Upper panel shows the scaled (with $4\pi r^2$) energy density $\varepsilon_{\rm slt}(r)$ and pressure $p_{\rm slt}(r)$ as functions of $r$. The lower panel shows the variation of $\varepsilon_{\rm slt}(r)$ and $p_{\rm slt}(r)$ as a function of r. The displayed quantities correspond to the soliton solution with the meson parameters of Ref.~\cite{Fukushima2020}.}
\label{fig:ener_pres_vs_r_Fukushima}
\end{figure}

The energy density $\varepsilon_{\rm{slt}}(r)$ and pressure $p_{\rm{slt}}(r)$ profiles of the soliton are shown in Fig.~\ref{fig:ener_pres_vs_r_Fukushima}. The pressure exhibits the characteristic feature of a positive core region and a negative surface region, ensuring global mechanical stability through the Virial theorem, $\int d^3 r\, p_{\rm{slt}}(r) = 0$. We find the soliton mass $M_{\rm{slt}} =\int \varepsilon_{\rm{slt}}(r) d^3 r  \approx 1452$~MeV, which agrees with the value $\sim 1460$~MeV reported in Ref.~\cite{Fukushima2020} within 0.5\%, confirming the accuracy of our numerical approach. An interesting quantity is the central ($r=0$) energy density and pressure, hereafter noted $\varepsilon^c_{\rm{slt}}$ and pressure $p^c_{\rm{slt}}$, which represent the point for which the energy density and the pressure are maximum as well as the point which is the further one from the surface. It is therefore the point that is the least contaminated by surface contributions.

In the context of the Skyrme model, it is well understood that the static soliton must be properly quantized in spin and isospin to reproduce the nucleon mass in its ground state. The static soliton represents a coherent superposition of spin–isospin states and therefore does not correspond to a pure nucleon state. A proper treatment would require collective quantization in spin and isospin, which generates the nucleon–Delta splitting through the moment of inertia and removes spurious rotational contributions~\footnote{ Though the quantization successfully explains the mass difference between nucleon ($N$) and Delta ($\Delta$)  baryon. The spin-isospin quantization increases the soliton mass by $3/8I$, $I$ being the moment of inertia. Furthermore, one needs to account for the center of mass correction for obtaining the mass of the nucleon. }. However, our main purpose is to address the in medium effect on the soliton EoS, which is barely discussed in ~\cite{Fukushima2020}. Furthermore  the quantization procedure is beyond the scope of the present work and can be addressed in the future work. Instead, we adopt the same mapping between the soliton model and nucleon as in ~\cite{Fukushima2020} and introduce a phenomenological scaling parameter $\chi$, defined at vacuum as,
\begin{equation}\label{eq:chi}
\chi= \frac{M_N^{\rm observed} \,(\sim 939~\mathrm{MeV})}{\left[M_{\rm{slt}}\right]_{\rm Vacuum}}\, .
\end{equation}
The scaling parameter ($\chi$)  should be viewed as a phenomenological correction for spurious collective effects inherent to the soliton, rather than as a substitute for full spin–isospin quantization. This scaling allows us to match the observed nucleon mass and enables a meaningful comparison between model predictions and physical observables. Accordingly, we rescale the soliton energy density and pressure profiles as follows~\cite{Fukushima2020}:
\begin{align}
\varepsilon_{\rm{slt},\chi}(r) &\equiv \chi \, \varepsilon_{\rm{slt}}(r)\,, \\
p_{\rm{slt},\chi}(r) &\equiv \frac{p_{\rm{slt}}(r)}{\chi}\,.
\end{align}

This rescaling procedure also facilitates direct comparison with the Nucleon EoS shown in Figure 8 of Ref.~\cite{Fukushima2020}. Based on their qualitative argument, a hard-core radius of $\sim 0.5$~fm suggests that hard deconfinement may set in around $\sim 8\rho_0$ (or possibly earlier). Furthermore, within the hard-confinement regime, it is argued that the thermodynamic pressure of bulk matter can be approximated by the mechanical pressure found at the core of the nucleon.

\begin{figure}[t]
\centering
\includegraphics[width=\linewidth]{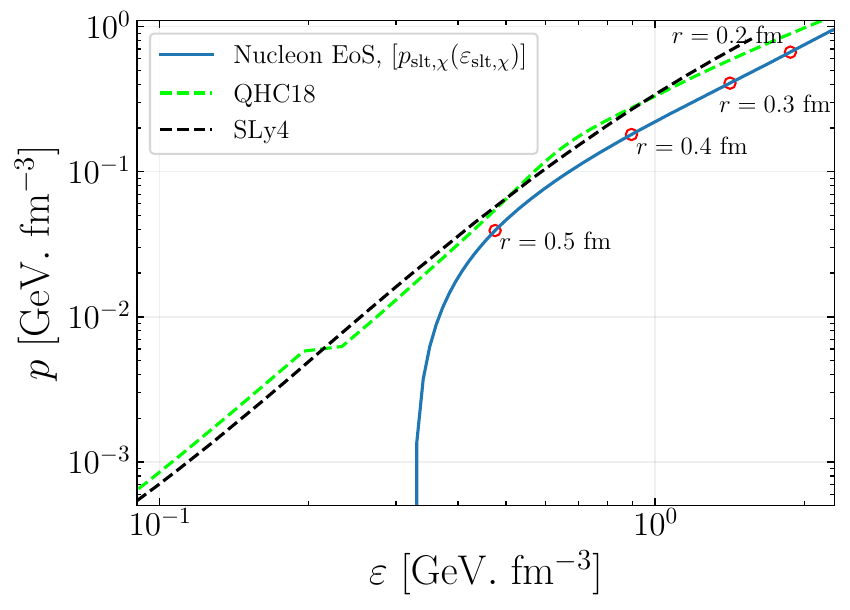}
\caption{Effective core EoS of the nucleon: pressure $p_{\rm slt,\chi}(r)$ as a function of energy density $\varepsilon_{\rm slt, \chi}(r)$, obtained by eliminating $r$ from the profiles shown in~Fig.~\ref{fig:ener_pres_vs_r_Fukushima}. Red circles indicate specific radial positions. For comparison, the solution~\cite{Gulminelli_2015_SLy4} of the SLy4 nuclear interaction~\cite{Chabanat:1998} and QHC18~\cite{Baym_QHC2018} EoS models are also shown.}
\label{fig:nucelon_EoS_Fukushima} 
\end{figure}

To isolate the surface effects specific to an isolated nucleon, we consider the pressure distribution near the center of the soliton~\cite{Fukushima2020}. As a test of our numerical implementation, we present the effective  EoS for the nucleon core—obtained by plotting $p_{\rm slt,\chi}(r)$ as a function of $\varepsilon_{\rm slt,\chi}(r)$ after eliminating the radial coordinate $r$. For context and comparison, we also include the NS $\beta$-equilibrated solution~\cite{Gulminelli_2015_SLy4} of the SLy4 nuclear interaction~\cite{Chabanat:1998} and the QHC18 model~\cite{Baym_QHC2018} obtained from the CompOSE~\cite{CompOSE2022} repository are also shown in  Fig.~\ref{fig:nucelon_EoS_Fukushima}.

\section{Results}\label{sec:results}
After successfully reproducing the key results from Ref.~\cite{Fukushima2020} with great accuracy, we now proceed to our objectives of this work: to investigate the sensitivity of vacuum soliton properties to the underlying NJL model parameters and to investigate the effect of chiral symmetry restoration in a self-consistent framework. The former analysis is presented in detail in section~\ref{sec:vacuum_soliton}. Furthermore, for the latter, we analyze the effects of chiral symmetry restoration on soliton structure and the corresponding  EoS, solely by means of the scalar field in section~\ref{sec:soliton_and_scalar_S} and in dense nuclear matter in section~\ref{sec:soliton_and_density}.

\subsection{NJL Parameters and Soliton in Vacuum}\label{sec:vacuum_soliton}
The sets of NJL parameters considered in this study are listed in Table~\ref{tab:NJL_parameters}. For completeness, we also listed the corresponding soliton model parameters evaluated at vacuum, such as the pion mass $M_\pi$, pion decay constant $f_\pi$, vector meson mass $M_v$, and effective couplings $g_v$ and $a$ in Table~\ref{tab:NJL_parameters}.

The parameter set labelled \texttt{NJL-F} is constructed to closely reproduce the vacuum soliton properties reported in Ref.~\cite{Fukushima2020}, but within the bosonization scheme adopted in this work, consistent with the formulation developed in Ref.~\cite{Chanfray2025}. The label ``F" in \texttt{NJL-F} refers to ``Fukushima et al.". In addition, we examine two further parameter sets inspired by the work of Ref.~\cite{Chanfray2025}. The set \texttt{NJL-C2} corresponds directly to the parameter set denoted as ``NJLSET2" in Ref.~\cite{Chanfray2025}, while \texttt{NJL-C1} represents an alternative choice within the same framework. Here, the label ``C" stands for ``Chanfray", and the number distinguishes between the different parameter choices. All these parameter sets are summarised in Table~\ref{tab:NJL_parameters}.

Although the \texttt{NJL-F} parametrization reproduces the meson properties (in particular the (iso-)vector sector),  reasonably well with the values used in Ref.~\cite{Fukushima2020}, we find that the alternative parametrization \texttt{NJL-FKW}, listed in Table~\ref{tab:NJL_parameters}, exactly reproduces both the scalar and vector meson properties. Consequently, the soliton EoS obtained with \texttt{NJL-FKW}  exactly reproduces the nucleon EoS of Ref.~\cite{Fukushima2020}, consistent with the results shown in Figs.~\ref{fig:Vacuum_EoS_scale_NJL}, and ~\ref{fig:Vacuum_EoS_noscale_NJL}. Despite this apparent advantage, we do not adopt \texttt{NJL-FKW} in our further analysis because the corresponding cut-off parameter is quite large, $\Lambda > 1~\text{GeV}$, along with the current quark mass, rather than being small $m_q\sim 1.5$ MeV. Nevertheless, the omission of this special parametrization does not affect the central motivation, main findings, or qualitative conclusions of the present work.

\begin{table*}[htbp]
    \centering
    \setlength\tabcolsep{.5em}
    \begin{tabular}{|c|c|c|c|c|c|c|c|c|c|c|c|}
    \hline\hline
    Name & \multicolumn{4}{c|}{NJL - Parameters} & \multicolumn{7}{c|}{Meson parameters at Vacuum} \\
    \cline{2-12}
         & $G_1$ & $G_2$ & $\Lambda$ & $m_q$ 
         & $M_0$ & $M_{\pi,\rm NJL}$ & $F_{\pi,\rm NJL}$ & $M_{\sigma}$ & $M_{v}$ & $g_v$ & $a$  \\
     & ($\rm GeV^{-2}$) &   ($\rm GeV^{-2}$) & (GeV) & ( MeV ) & ( MeV ) & ( MeV ) &  ( MeV ) & ( MeV ) & ( MeV ) & &\\
     \hline
     NJL-F & 4.3 & 14.74 & 0.950 &  1.95 & 331.874 & 140.378 & 87.392 & 906.180 & 818.455 &3.142 & 2.221\\
     \hline     
     NJL-C1 & 7.5 & 7.5 & 0.735 & 3.6 & 308.089 & 139.553 & 87.140&714.951 & 1253.316 & 3.432& 4.390  \\
     \hline
     NJL-C2 & 10 & 7.5 & 0.665 & 4 & 358.840 & 138.185 &86.700 & 827.123 & 1447.195 & 3.963 & 4.434\\
     \hline
     NJL-FKW & 3.079 & 14.679 & 1.11 & 1.47 & 348.881 & 140 &92.277 & 996.667 & 783 & 3 & 2\\
     \hline
    \hline\hline
    \end{tabular}
    \caption{NJL-Model parameters and corresponding scalar and  (iso-) vector  properties at vacuum (${\cal S}=M_0$ or $s=0$).}
    \label{tab:NJL_parameters}
    \end{table*}    
  
\begin{figure}[htbp]
     \centering
     \includegraphics[width=0.95\linewidth]{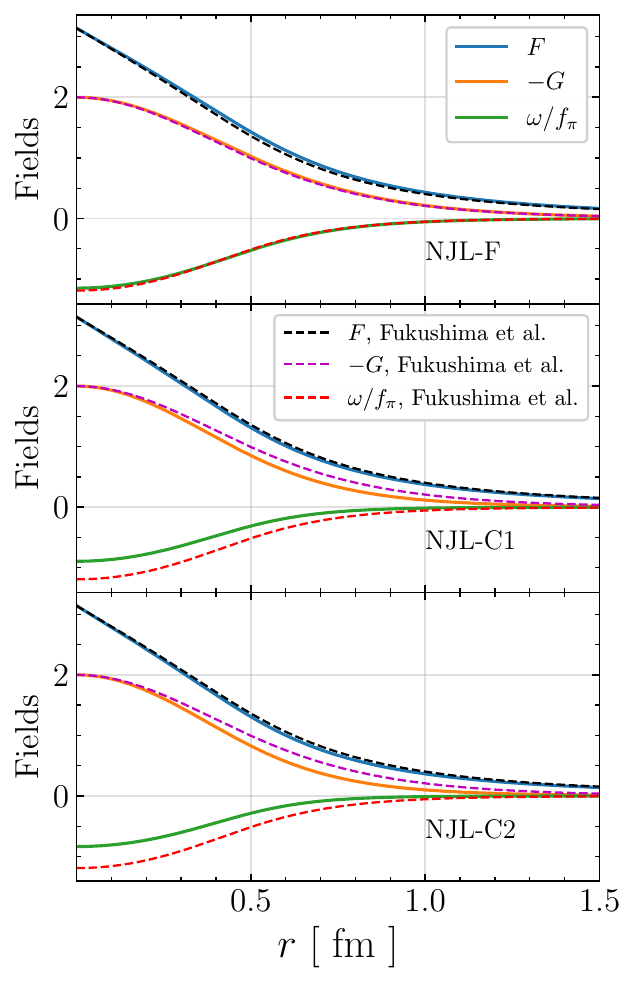}
     \caption{Solid lines in different panels show the meson field profiles for different NJL parameters as noted in the sub-panel. For comparison, the field profiles corresponding to the parameters of Ref.~\cite{Fukushima2020} are also shown in dashed curves.}
     \label{fig:Fields_NJL_Parameter_Vacuum}
 \end{figure} 
 \begin{figure}[htbp]
     \centering
     \includegraphics[width=\linewidth]{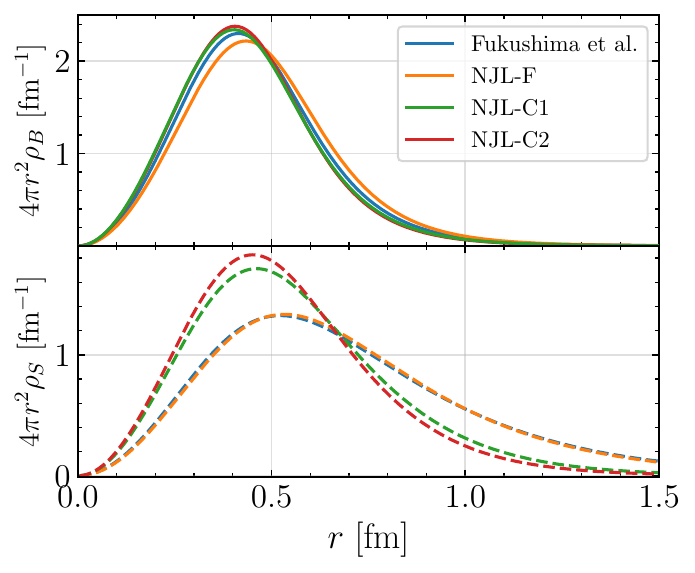}
     \caption{Upper (lower) panel: scaled baryon (isoscalar) charge density as a function of $r$.  The same color denotes the same NJL parametrization in both panels.}
     \label{fig:Density_NJL_Parameter_Vacuum}
 \end{figure}
 
The meson field profiles for the \texttt{NJL-F}, \texttt{NJL-C1}, and \texttt{NJL-C2} parameter sets are shown in Fig.~\ref{fig:Fields_NJL_Parameter_Vacuum}, alongside the profiles corresponding to the parameters from Ref.~\cite{Fukushima2020} for comparison. Key soliton properties such as soliton mass $M_{\rm slt}$, baryon charge radius $\sqrt{\langle r_B^2 \rangle}$, and isoscalar charge radius $\sqrt{\langle r_S^2 \rangle}$ at vacuum are summarized in Table~\ref{tab:Soliton_properties_in_vacuum}. As the vector meson mass $M_v$ increases across the parameter sets, the $\omega$-field becomes more localized near the core of the soliton, as evident in Fig.~\ref{fig:Fields_NJL_Parameter_Vacuum}. This localization leads to a sharper radial decay, resulting in a reduced isoscalar charge radius $\sqrt{\langle r_S^2 \rangle}$, as shown in Fig.~\ref{fig:Density_NJL_Parameter_Vacuum} and quantified in Table~\ref{tab:Soliton_properties_in_vacuum}. In contrast, the baryon charge radius $\sqrt{\langle r_B^2 \rangle}$, associated with the valence quark distribution, remains relatively insensitive to changes in $M_v$, indicating the robustness of the hard-core structure of the soliton.
 
\begin{table}[htbp]
\centering
        \begin{tabular}{|c| c|c|c| c| c|}
        \hline \hline
       Name& $M_{\rm slt}$ & $\sqrt{<r_B^2>}$ & $\sqrt{<r_S^2>}$ & $g_S$ & $C_{\rm NS}$ \\
        & $\rm (\ MeV\ )$ & ($\rm fm$) & ($\rm fm$) &  &\\
        \hline
        NJL-F&   1350 & 0.514 & 0.778 &  2.64 & -0.16\\
        \hline
        NJL-C1 & 1216 & 0.484 & 0.621 & 2.22  &-0.24 \\
        \hline
         NJL-C2 & 1137 & 0.481 & 0.587 & 1.30 &-0.19 \\        
        \hline \hline
        \end{tabular} 
        \caption{Soliton properties evaluated at vacuum for different NJL parameter sets.}
        \label{tab:Soliton_properties_in_vacuum}
\end{table}

We also present the nucleon-scalar coupling $g_S$ and the scalar susceptibility $C_{\rm NS}$  within the NJL-soliton framework, defined as in Refs.~\cite{Chanfray2024EPJA,Chanfray2025}:
\begin{align}
g_S&=\frac{M_0}{f_{\pi}}\left(\frac{\partial M_N}{\partial S}\right)_{S=M_0} ~,\\
C_{NS}&=\frac{M_0^2}{2M_N}\left(\frac{\partial^2 M_N}{\partial {S}^2}\right)_{S=M_0} ~.
\end{align}

We observe a notably small nucleon-scalar coupling\footnote{Alternatively, if one evolve the soliton only through the scalar sector of the NJL-soliton model ($F_{\pi}({\cal S}), M_{\pi}({\cal S})$), while keeping the vector parameters fixed at their vacuum values [$M_v({\cal S})=M_v(M_0),\ g_v({\cal S})=g_v(M_0)$], as done in several works, see Refs.~\cite{CHRISTOV1990,Alkofer1991}, leads to a different nucleon scalar coupling, as discussed in ~\ref{app:gs_and_mv}.
} constant within the present framework, accompanied by a negative scalar susceptibility. The appearance of a negative susceptibility is consistent within the  {NJL-Soliton/skyrmion model}. However, we anticipate that incorporating confinement effects, {for instance within the chiral confining model}, will mitigate this behavior, potentially yielding a positive susceptibility more in line with the results reported in Refs.~\cite{Chanfray2024EPJA,Chanfray2025}.
\begin{figure*}[htbp]
    \centering
    \begin{subfigure}{0.49\textwidth}
    \includegraphics[width=\linewidth]{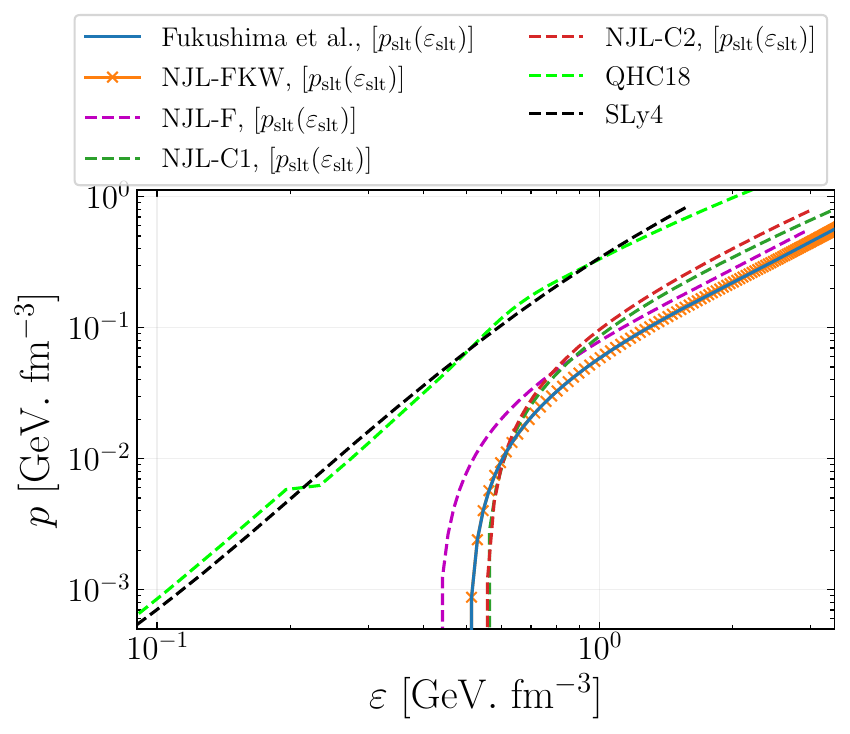}
    \caption{}
    \label{fig:Vacuum_EoS_noscale_NJL}
    \end{subfigure}
    \begin{subfigure}{0.49\textwidth}
    \includegraphics[width=\linewidth]{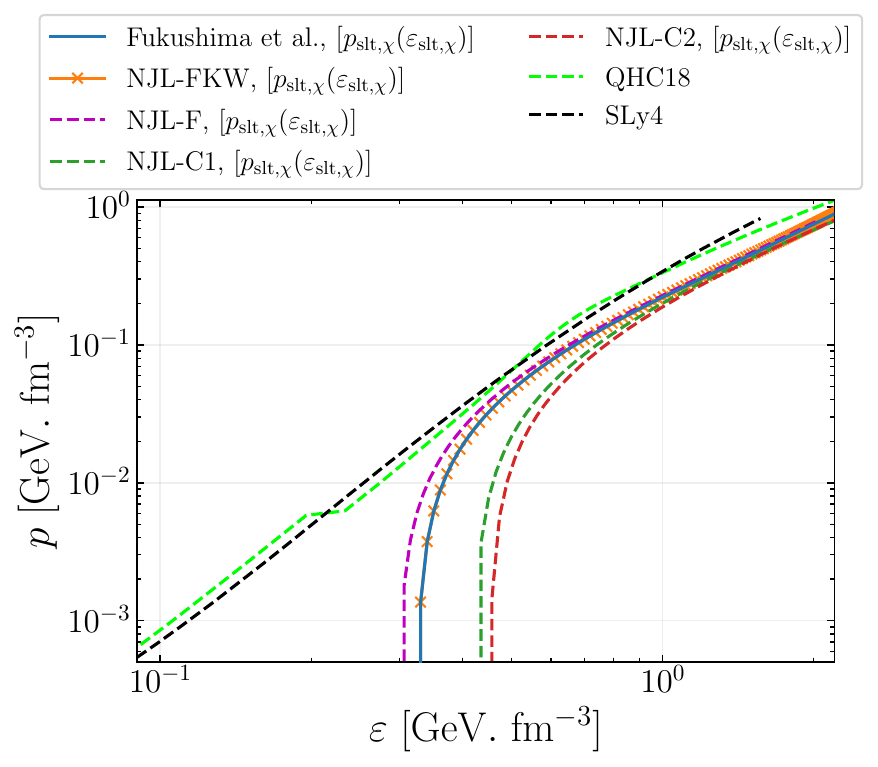}
    \caption{}
    \label{fig:Vacuum_EoS_scale_NJL}
    \end{subfigure}    
   \caption{Soliton pressures as a function of energy density for different NJL parameter sets, with $p_{\rm slt}(r)$–$\varepsilon_{\rm slt}(r)$ shown in the (a) and $p_{\rm slt,\chi}(r)$–$\varepsilon_{\rm slt,\chi}(r)$ in the (b). The QHC18 and SLy4 EoSs are included for comparison.}
\end{figure*}
Although we adopt a normalization procedure involving the energy density, specifically multiplying by the factor $\chi$, to reproduce the physical nucleon mass at vacuum (as outlined in section~\ref{sec:Numerical_method}), we present and compare the resulting EoS both with and without this approach. The EoSs for the various NJL parameter sets, obtained by eliminating the radial coordinate $r$ from the profiles of energy density $\varepsilon_{\rm slt}(r)$ and pressure $p_{\rm slt}(r)$, are shown in Fig.~\ref{fig:Vacuum_EoS_noscale_NJL}, alongside the reference EoS corresponding to the exact parameters of Ref.~\cite{Fukushima2020}. The impact of the scaling parameter $\chi$ on the EoS is illustrated in Fig.~\ref{fig:Vacuum_EoS_scale_NJL}, and the scaling approach changes the stiffness of the EoSs.

\subsection{Effects of Chiral Symmetry Restoration}

\subsubsection{Scalar field dependency}
\label{sec:soliton_and_scalar_S}

In our model framework, the explicit effects of chiral symmetry breaking enter the soliton description solely through the scalar field $\cal S$. Before extending the discussion to dense matter and analyzing how both the scalar field and soliton evolve in such environments, we begin by exploring the isolated effect of chiral symmetry breaking on soliton properties as governed by the scalar field $\cal S$.

To this end, we artificially vary the nuclear scalar field $s$, and compute the corresponding scalar field $\cal S$ using the relation provided in Eq.~\eqref{eq:nuclear_s_to_S}. Given a fixed set of NJL parameters ($\{G_1, G_2, m_q, \Lambda\}$) and the value of $\cal S$, we calculate the in-medium quantities: the pion decay constant $F_\pi({\cal S})$, pion mass $M_\pi({\cal S})$, vector meson mass $M_v({\cal S})$, vector coupling $g_v({\cal S})$, and the parameter $a({\cal S})$. With these medium-modified quantities, we solve for the soliton configuration following the numerical procedure described in section~\ref{sec:Numerical_method}.

\begin{figure*}[htbp]
\centering
\includegraphics[width=\linewidth]{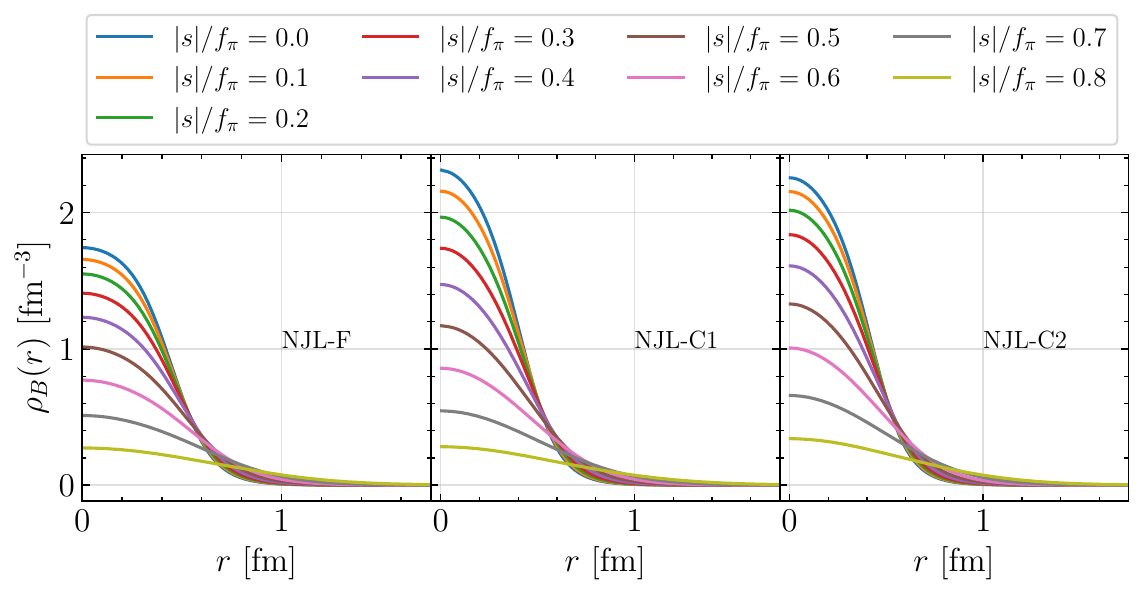}
\caption{$\rho_B(r)$ versus $r$ obtained for different $s$ values, with different NJL parameterizations as labeled in the sub-panels.}
\label{fig:nb_vs_r_diff_s}
\end{figure*}

\begin{figure}[t]
\centering
\includegraphics[width=\linewidth]{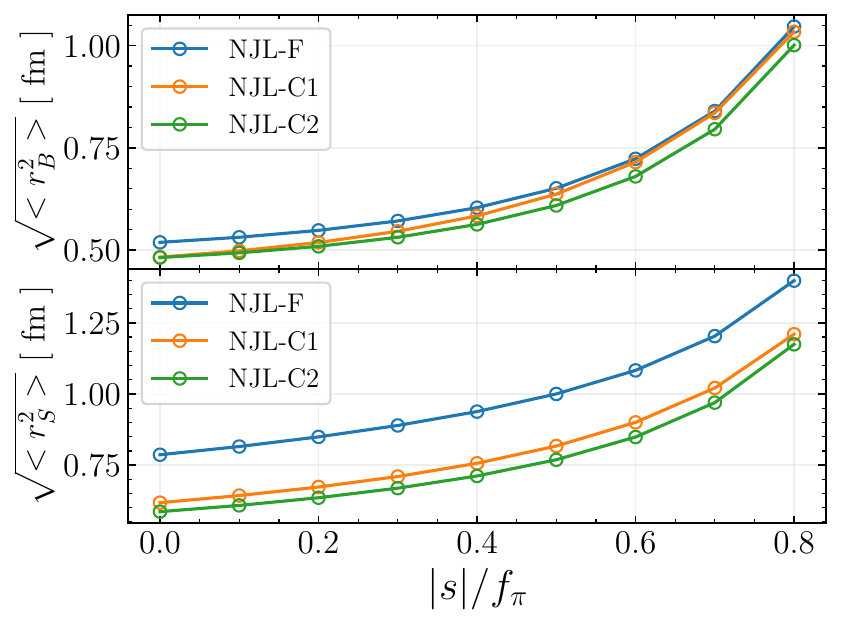}
\caption{ In upper (lower) panel we show the variation of $\sqrt{<r_B^2>}$ ( $\sqrt{<r_{S}^2>}$) as a function of $s$~.} 
\label{fig:rb_rs_vs_s} 
\end{figure}

The local baryon density ($\rho_B(r)$) profiles (carried by valence quarks) of the soliton for various values of $s$ are shown in Fig.~\ref{fig:nb_vs_r_diff_s}. As the scalar field strength increases, which corresponds to more restored chiral symmetry (or higher matter density as discussed in section~\ref{sec:soliton_and_density}), we observe that the quarks become progressively delocalized. This results in the baryon density spreading over a larger spatial region, effectively increasing the hard-core size of the nucleon as displayed in Fig.~\ref{fig:rb_rs_vs_s}. Simultaneously, the central density decreases due to the reduced effective spatial quark confinement. This behavior is also reflected in the energy density profiles, as illustrated in~Fig.~\ref{fig:EoS_vs_s}.

For sufficiently large scalar fields $s$, the notion of a localized hard core begins to break down altogether. The evolution of the baryon charge radius (interpreted as the hard-core radius) and the isoscalar charge radius with respect to $s$ is presented in Figs.~\ref{fig:rb_rs_vs_s}, \ref{fig:rb_rs_vs_s}, respectively. The in-medium effect on the size of the hard core, which increases with density, indicates that the available volume for the nucleons in a nuclear medium will decrease, and the overlap of the hard cores begins at a density smaller than that corresponding to the overlap density calculated from the vacuum baryon charge radius. 

\begin{figure}[htbp]
\centering
  \includegraphics[width=\linewidth]{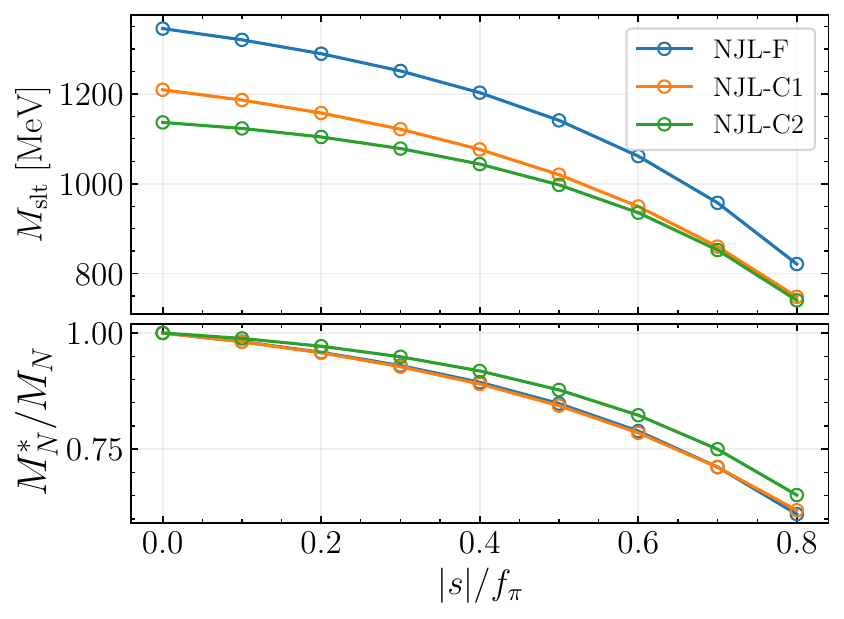}
  \caption{ Upper  panel: variation of $M_{\rm slt}$ as a function of $s$. Lower panel: $M^*_{N}(s)/M_{N}(s=0)$ as a function of $s$.}\label{fig:Mn_vs_s} 
\end{figure}

\begin{figure*}[htbp]
\centering
\includegraphics[width=\linewidth]{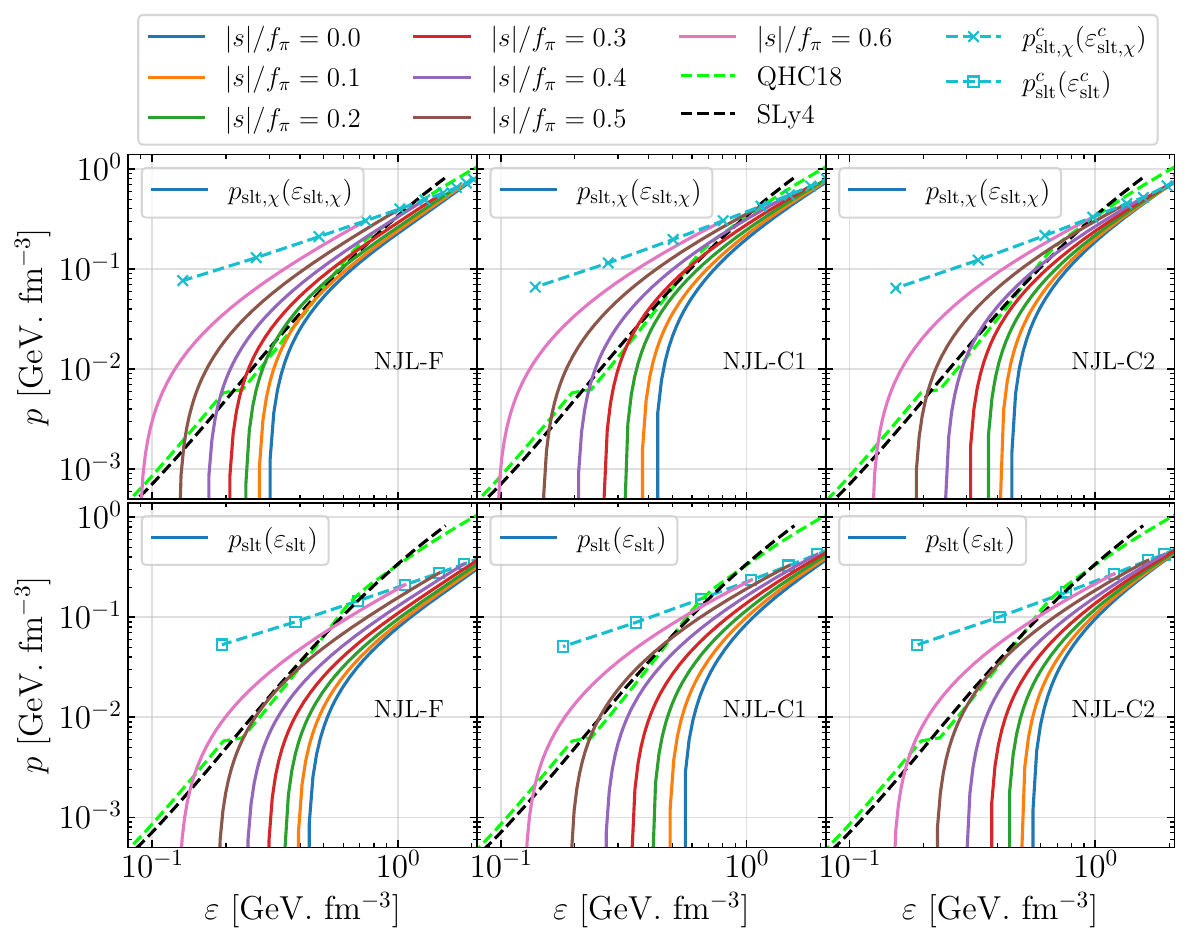}
\caption{Upper panels: $p_{\rm slt,\chi}(r)$ versus $\varepsilon_{\rm slt,\chi}(r)$ for different $s$ values and NJL parameterizations (labeled in the sub-panels). Lower panels: same as above, but for $p_{\rm slt}(r)$ versus $\varepsilon_{\rm slt}(r)$. For comparison we have also displayed the evolution of central pressure as a function of the central energy density. In all sub-panels, NS EoSs (SLy4 and QHC18) are shown with dashed lines, while soliton-based EoSs are shown with solid lines.}
\label{fig:EoS_vs_s}
\end{figure*}

To ensure consistency with the observed nucleon mass $M_N$ in vacuum ($s=0$), we initially scale the energy density $\varepsilon_{\rm slt}(r)$ by a normalization constant $\chi$  defined in Eq.~\eqref{eq:chi}. For non-zero scalar field values corresponding to in-medium conditions, we retain the same scaling factor $\chi$ determined in vacuum. This ensures that the comparison across different values of $s$ reflects in-medium modifications to the nucleon mass $M^*_N=\chi M_{\rm slt}(s)$. Hence, we re-normalized the EoSs at each $s$ by the scaling factor $\chi$ evaluated in vacuum. The resulting in-medium soliton mass $M_{\rm slt}(s)$ as a function of the scalar field is shown in Fig.~\ref{fig:Mn_vs_s}, which demonstrates the continuous reduction of the soliton (nucleon) mass as chiral symmetry is progressively restored.

The local soliton EoSs (both $p_{\rm{slt},\chi}(\varepsilon_{\rm{slt},\chi})$ and $p_{\rm{slt}}(\varepsilon_{\rm{slt}})$), obtained for various values of the scalar field $s$ are shown in Fig.~\ref{fig:EoS_vs_s}. In contrast to the argument of Ref.~\cite{Fukushima2020},  where the effect of chiral symmetry restoration on the nucleon EoS was taken into account by rescaling the energy density and pressure by an in-medium pion decay constant $f^*_{\pi}$, to this end, we have demonstrated the effect of chiral symmetry restoration on the interior EoS of the soliton in a self-consistent manner. As the strength of the scalar field $s$ (or density) increases and reaches to $0.2 f_{\pi}-0.3 f_{\pi}$ ({or equivalently with the in-medium pion decay constant ($F_{\pi}^*$) approaches 0.85-0.93 times the vacuum value $f_\pi$} as shown in ~~\ref{app:mesons_in_nuclear_matter}), the re-normalized (by the factor $\chi$) EoS reaches a stiffness compared to that of the other preferred models of the high density NS matter (as displayed for SLy4 and QHC18). Whereas, ignoring the scaling of the EoS by $\chi$, one would reach the stiffness of the NS EoSs roughly for $s\sim 0.5f_{\pi}$ {equivalently $F_{\pi}^*\to 0.7f_{\pi}-0.75 f_{\pi} $}). 

\begin{figure*}[t]
\centering
\begin{subfigure}{0.49\textwidth}
\includegraphics[width=\linewidth]{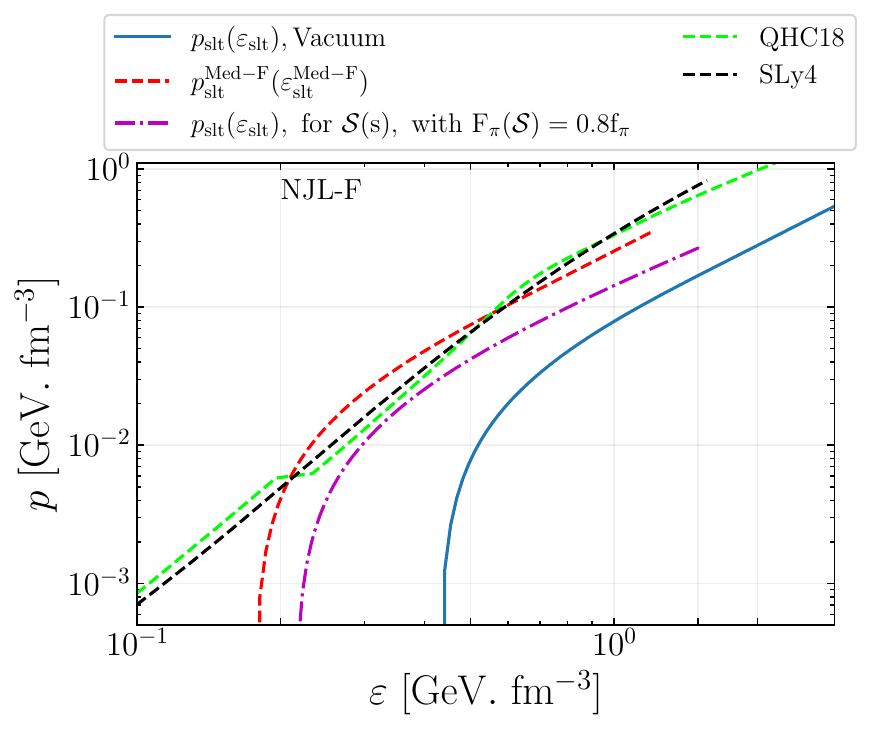}
\caption{ }
\label{fig:EoS_and_Fpi}
\end{subfigure}
\begin{subfigure}{0.49\textwidth}
\includegraphics[width=\linewidth]{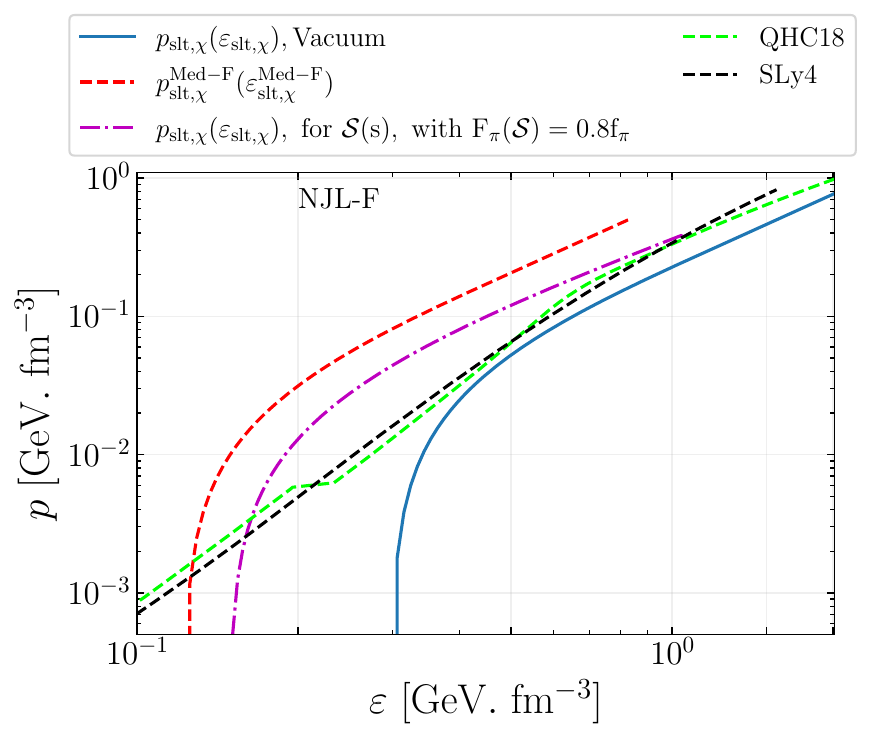}
\caption{}
\label{fig:scaled_EoS_and_Fpi}
\end{subfigure}    
\caption{(a) Soliton pressure $p_{\rm slt}(r)$ versus energy density $\varepsilon_{\rm slt}(r)$ for the NJL-F parametrization. The vacuum EoS (blue solid line) is compared with the in-medium modified soliton EoS (red dashed line) obtained using Eqs.~\eqref{eq:e_medium_Fukushima}-\eqref{eq:p_medium_Fukushima} with $F^*_{\pi}=0.8f_{\pi}$, and with the self-consistent soliton EoS (dot-dashed line) from section~\ref{sec:Method_NJL} for a scalar field $\mathcal{S}$ such that $F^*_{\pi}(\mathcal{S})=0.8f_{\pi}$. (b) Same as (a), but for $p_{\rm slt,\chi}(r)$ versus $\varepsilon_{\rm slt,\chi}(r)$, with $\chi$ fixed to its vacuum value.}
    \label{fig:Inmedium_EoS_and_Fpi}
\end{figure*}

Following the argument of Ref.~\cite{Fukushima2020}, the in-medium soliton energy density $\varepsilon^{\rm Med-F}_{\rm slt}$ and pressure $p^{\rm Med-F}_{\rm slt}$ can be obtained by rescaling their vacuum counterparts as
\begin{align} 
\varepsilon^{\rm Med-F}_{\rm slt}(r)\;
\equiv
\;\left(\frac{F_\pi^*}{f_\pi}\right)^4 \left[\varepsilon_{\rm slt}(r)\right]_{\rm Vacuum}, \label{eq:e_medium_Fukushima} \\
\qquad p^{\rm Med-F}_{\rm slt}(r)\;
\equiv
\;\left(\frac{F_\pi^*}{f_\pi}\right)^2 \left[p_{\rm slt}(r)\right]_{\rm Vacuum} \label{eq:p_medium_Fukushima} 
\end{align}
where $F^*_{\pi}$ and $f_\pi$ denote the in-medium and vacuum pion decay constants, respectively. To examine this argument and compare it with results from our framework, we present in-medium EoSs for the NJL-F parametrization obtained in different approaches in Fig.~\ref{fig:Inmedium_EoS_and_Fpi} as follows:
\begin{itemize}
\item For clarity, results without and with normalization of EoS by $\chi$ (note that $\chi$ is fixed to its vacuum value) are displayed in Fig.~\ref{fig:EoS_and_Fpi} and Fig.~\ref{fig:scaled_EoS_and_Fpi}, respectively~.
\item In Fig.~\ref{fig:Inmedium_EoS_and_Fpi}, the vacuum soliton EoSs is shown by solid lines. 
\item The in-medium soliton EoS for $F_\pi^* = 0.8 f_\pi$, obtained using Eqs.~\eqref{eq:e_medium_Fukushima} and \eqref{eq:p_medium_Fukushima} (following Ref.~\cite{Fukushima2020}), is shown by dashed lines in Figs.~\ref{fig:EoS_and_Fpi} and ~\ref{fig:scaled_EoS_and_Fpi}, respectively.
\item For comparison, in Figs.~\ref{fig:EoS_and_Fpi}, and ~\ref{fig:scaled_EoS_and_Fpi}, we also display the soliton EoS obtained in our framework for a scalar field ${\cal S}(s)$ such that $F_\pi^*({\cal S}) = 0.8 f_\pi$ (see section~\ref{sec:Method_NJL}), shown by dot–dashed lines.
\end{itemize}

We notice from Fig.~\ref{fig:Inmedium_EoS_and_Fpi},  that if one considers the rescaling approach of Ref.~\cite{Fukushima2020}, as discussed in Eqs.~\eqref{eq:e_medium_Fukushima} and \eqref{eq:p_medium_Fukushima}, it tends to overestimate the stiffness of the EoS compared to the fully self-consistent approach adopted in this work. Qualitatively, this difference can be understood as a result of the competing effects from the scalar and vector sectors. The decrease of $F_\pi$ with increasing $s$ tends to stiffen the EoS, while the simultaneous reduction of $M_v$ acts in the opposite direction, softening the EoS in the core region. This highlights that a simple rescaling prescription cannot capture the interplay between scalar and vector contributions.  

\subsubsection{ Density dependence }
\label{sec:soliton_and_density}

In this section, we extend our investigation on the in-medium modification of soliton properties and the EoS induced by the restoration of chiral symmetry relating to dense nuclear matter. In our framework, we assume that quarks describe the vacuum dynamics through spontaneous chiral symmetry breaking and mesonic degrees of freedom, while nuclear matter itself is treated as a collection of nucleons interacting via the meson fields generated by the underlying quark dynamics. The nucleons (baryon number $1$ solitons) are therefore the relevant degrees of freedom for matter. In the present work, however, we neglect such corrections and consider instead that, within nuclear matter, quarks do not propagate explicitly and the Dirac sea contribution is subtracted, as it naturally appears in the chiral scalar potential of the underlying NJL model in its bosonized form. Though the three-momentum cut-off on the underlying NJL model is not covariant, its use is standard and  commonly adopted in NJL-based approaches~\cite{Klevansky1992,Fukushima2013}. A more microscopic description, based on solving the quark wavefunctions in the soliton background, could provide direct access to the finite spatial extent and thus allow a more consistent treatment of finite-size effects.  For simplicity, we further restrict our study to symmetric nuclear matter (SNM) and focus on the evolution of soliton properties in this environment. 

The energy density of nuclear matter, $\varepsilon_N$, at nuclear matter density ($\rho_N$) in the mean-field approximation is given by~\cite{Chanfray2010b}:
\begin{equation}\label{eq:energy_density_nuclear}
\varepsilon_N = \frac{4}{(2\pi)^3}\int_0^{k_F} d^3k \
\sqrt{k^2+M^{*2}_N(s)}
+ V(s) + \frac{9}{2}G_2 \rho_N^2,
\end{equation}
where the nuclear Fermi momentum $k_F$ is related to the nuclear matter density $\rho_N$ via
\begin{equation}
\rho_N =  \frac{2}{3\pi^2}k_F^3.
\end{equation}

The in-medium, scalar-field–dependent effective nucleon mass is provided by the NJL–soliton model (see Sections~\ref{sec:NJL_model} and \ref{sec:soliton_and_scalar_S}) and is expressed as
\begin{equation}
M^*_N(s) = \chi \, M_{\rm slt}(s).
\end{equation}

Subtracting the vacuum contribution, the chiral effective potential takes the form
\begin{align}\label{eq:NJL_chiral_potential}
V({\cal S}) 
= &-2N_cN_f\big[I_0({\cal S})-I_0(M_0)\big] \nonumber\\
&+ \frac{({\cal S}-m_q)^2-(M_0-m_q)^2}{2G_1},
\end{align}
where the first term represents the (in-medium) energy of the Dirac sea of constituent quarks with
\begin{equation}
I_0({\cal S})=\int_0^{\Lambda} \frac{d^3p}{(2\pi)^3} \sqrt{p^2+{\cal S}^2}.
\end{equation}

In the mean-field limit, the in-medium scalar field $s$ is obtained by minimizing the energy density Eq.~\eqref{eq:energy_density_nuclear} with respect to $s$. This yields the equation of motion for the scalar field `$s$':
\begin{equation}\label{eq:nuclear_s_field}
\frac{\partial V}{\partial s}=-\frac{\partial M_N^{*}}{\partial s} \, n_S \equiv -g^*_s n_S,
\end{equation}
where the effective scalar coupling is defined as
\begin{equation}\label{eq:g_s_density}
g^*_s=\frac{\partial M_N^{*}}{\partial s},
\end{equation}
and the scalar density of SNM is
\begin{equation}
n_S=\frac{\gamma}{(2\pi)^3}\int_0^{k_F} d^3k \,
\frac{M^{*}_N(s)}{\sqrt{k^2+M^{*2}_N(s)}}\ , \quad \gamma=4~.
\end{equation}

Although our starting formulation is  self consistent embedding the $s$-dependent soliton mass obtained from the NJL-soliton model directly into the mean-field energy density—we find that, when the density-dependent scalar coupling $g_s^*(s)$ (defined in Eq.~\eqref{eq:g_s_density}) is considered, the solution of the in-medium gap equation Eq.~\eqref{eq:nuclear_s_field} ceases to exist beyond $\rho_N \sim 3\rho_{\rm sat}$. This breakdown originates from the rapid growth of $g_s^*(s)$ with increasing density, which causes the source term $-g_s^*(s)\,n_S$ to increase too strongly to be balanced by the scalar potential term $\partial V/\partial s$. As a consequence, the scalar field equation no longer admits a physical solution, preventing a meaningful exploration of soliton properties at higher densities within this fully self consistent setup.  To enable exploration of soliton properties at higher densities, we therefore follow a common practice in RMF approaches and fix the scalar coupling to its vacuum value, as given in Eq.~\eqref{eq:gs_fix}.

\begin{equation}\label{eq:gs_fix}
g_s^*=\left.\frac{\partial M_N^{*}}{\partial s}\right|_{s=0}.
\end{equation}
\begin{figure}
    \centering
    \includegraphics[width=\linewidth]{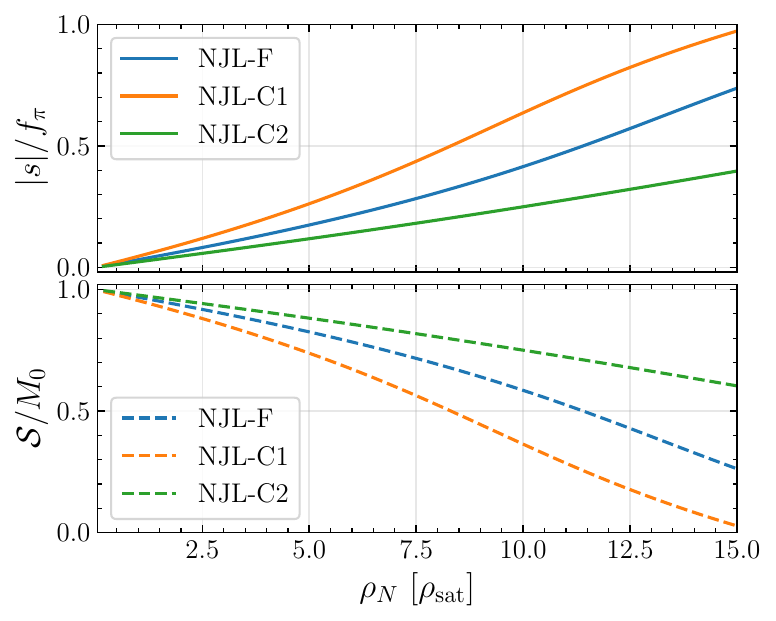}
    \caption{Upper panel: The density dependency of the nuclear scalar field ($s$) as a function of the density for different NJL parameters considered in this work. The solutions are obtained by solving Eq.~\eqref{eq:nuclear_s_field} with fixing  $g_s^*$ from ~Eq.~\eqref{eq:gs_fix}. Lower Panel: same as upper panel but for the scalar field $\cal S$, obtained using Eq.~\eqref{eq:nuclear_s_to_S}. }
    \label{fig:s_vs_rho}
\end{figure}
Physically, confinement effects are also expected to moderate the abrupt drop of the soliton mass with $s$, supporting this approximation. Thus, for a given density $\rho_N$, the corresponding nuclear scalar field $s({\cal S})$ is obtained from the solution of Eq.~\eqref{eq:nuclear_s_field} and the density evolution of ${\cal S}$ is  displayed in Fig.~\ref{fig:s_vs_rho}. Then at a given ${\cal S}(\rho_N)$ the soliton properties are computed following section~\ref{sec:NJL_model}. In this way, we investigate the impact of in-medium chiral symmetry restoration on soliton properties and the nuclear EoS.  

\begin{figure}[t]
\centering
\includegraphics[width=0.98\linewidth]{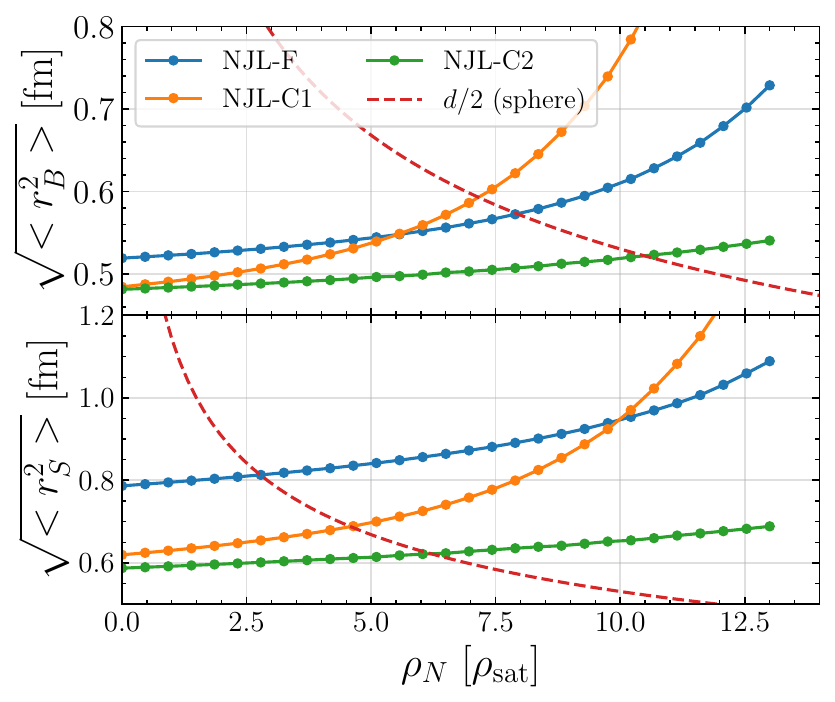}
\caption{Evolution of the baryon charge radius $\sqrt{\langle r_B^2\rangle}$ (upper panel) and scalar charge radius $\sqrt{\langle r_S^2\rangle}$ (lower panel) as functions of $\rho_N$. The half inter-nucleon distance $d/2$ in uniform matter is also shown for reference.}
\label{fig:rb_rs_vs_rhoN}
\end{figure}

The density evolution of the baryon and scalar charge radii is shown in~Fig.~\ref{fig:rb_rs_vs_rhoN}, together with the half inter-nucleon distance $d/2\sim \left(4\pi/3\rho_N\right)^{-1/3}$ for reference. As shown in  Fig.~\ref{fig:rb_rs_vs_rhoN}, the soliton hard cores overlap once the baryon charge radius approaches $d/2$, which occurs at  $\rho_N \sim 8\rho_{\rm sat},\ 7\rho_{\rm sat}, \ \text{and } 10\rho_{\rm sat}$ for NJL-F, NJL-C1, and NJL-C2 parameter sets, respectively. This indicates the onset of hard deconfinement and the possible emergence of quark matter. In contrast, the overlap of the soft cores, which enhances quark mobility, takes place already at lower densities, around $\rho_N \sim 3\rho_{\rm sat},\ 4\rho_{\rm sat}, \ \text{and } 6\rho_{\rm sat}$ for NJL-F, NJL-C1, and NJL-C2, respectively. A redefinition of the inter-nucleon spacing $d/2$ under alternative close-packing assumptions would further lower these thresholds, suggesting that deconfinement may set in at densities below the quoted values.

To construct a density-dependent EoS from the chiral soliton ansatz, we consider the matter as a juxtaposition of quark cores ($QC$), each characterized by a hard-core radius of order $\sqrt{\langle r_B^2 \rangle}$. At bulk baryon density $\rho_N$, the number of quark cores per unit volume is proportional to $\rho_N$, such that the quarks are adjusted to the average bulk energy density $\varepsilon_{QC} (\rho_N)$ given as:
\begin{equation}
\varepsilon_{QC} (\rho_N) \equiv  M_{\rm slt}(s,s(\rho_N))\, \rho_N~.
\end{equation}
While the energy density inside each soliton core has a spatial profile, we implicitly assume that, on average, the quarks experience an effective energy density equal to $\varepsilon_{QC}$. Thus, at a given baryon density $\rho_N$ with the corresponding nuclear scalar field $s(\rho_N)$, we determine the radial distance $r=r_{QC}$ from the center of a soliton core (obtained for the given $s(\rho_N)$ following section~\ref{sec:NJL_model}) at which the local energy density  {$\varepsilon_{\rm slt}(r_{QC},s(\rho_N))$} matches the bulk value $\varepsilon_{QC}$. That is,
\begin{equation}
\varepsilon_{QC}(\rho_N)=\varepsilon_{\rm slt}(r_{QC},s(\rho_N))~.
\end{equation}
At this distance $r=r_{QC}$, we extract the corresponding pressure from the soliton's pressure profile $p_{\rm slt}(r)$ as,
\begin{equation}\label{eq:EoS_Eqc}
p(\varepsilon_{QC})=p_{\rm slt}(r,s(\rho_N) )\Big |_{r=r_{QC}} ~.
\end{equation}

\begin{figure}[t]
\centering
\includegraphics[width=\linewidth]{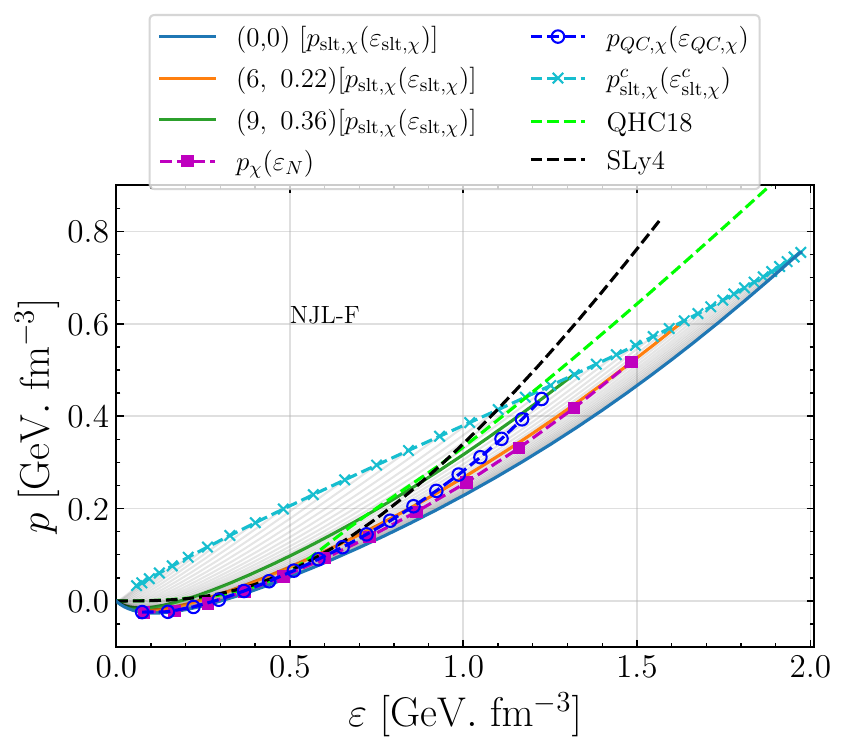}
\caption{Soliton pressure $p_{\rm slt,\chi}(r)$ versus energy density $\varepsilon_{\rm slt, \chi}(r)$ at various nuclear matter densities and corresponding scalar fields. The legend indicates $\rho_N$ and $s$ as $(\rho_N/\rho_{\rm sat},\, |s|/f_{\pi})$. Results are shown for the NJL-F parameter set. For reference, the evolution of the central pressure $p_{\rm slt}^c$ with central energy density $\varepsilon_{\rm slt}^c$ is also displayed. To improve visibility, soliton EoSs at selected densities are highlighted in color and labeled, while the others are shown in black. The EoS $p_{QC,\chi}(\varepsilon_{ QC,\chi})$ is displayed with dashed circled line.  For comparison, the resulting EoS with ${p_\chi}(\varepsilon_N)$ is also displayed.  NS EoSs (SLy4 and QHC18) are also shown for comparison.}
\label{fig:EoS_Eqc_NJLF}
\end{figure}

In this way, we map the soliton energy-density profile to the bulk average density to obtain the EoS. To elucidate the density-dependent EoS derived from the chiral soliton ansatz, the resulting EoS along with the procedure is illustrated in Fig.~\ref{fig:EoS_Eqc_NJLF} for the parameter set NJL-F. For clarity, EoSs corresponding to only selected nuclear matter densities $\rho_N$ are displayed in the figures. Our analysis reveals that \(\varepsilon_\text{QC}\) increases  with \(\rho_N\), consistent with a denser packing of QCs. However, the maximum local energy density within the soliton core (\(\varepsilon(r)\)) decreases as the in-medium scalar field \({\cal S}(s)\) diminishes with increasing density, signaling partial chiral restoration. This competition naturally defines a breakdown density $\rho_N$ (or equivalently a limiting $\varepsilon_{QC}$), beyond which the soliton-based EoS ceases. Specifically, the density-dependent EoS truncates at critical densities of approximately \(\rho_N \sim 9 \rho_\text{sat}\), \(8 \rho_\text{sat}\), and \(12 \rho_\text{sat}\) for the NJL-F, NJL-C1, and NJL-C2 parameter sets, respectively. These termination points can be interpreted as critical densities at which quarks are no longer confined within individual solitons but instead percolate into a collective deconfined phase of quark matter. Additionally, the onset of this breakdown coincides with the densities at which the soliton hard cores begin to overlap. This truncation suggests a physical limit to the soliton-based description, analogous to a phase transition where quarks are no longer confined within distinct solitons but form a dense, homogeneous quark matter.

\begin{figure*}[htbp]
    \centering
    \includegraphics[width=\linewidth]{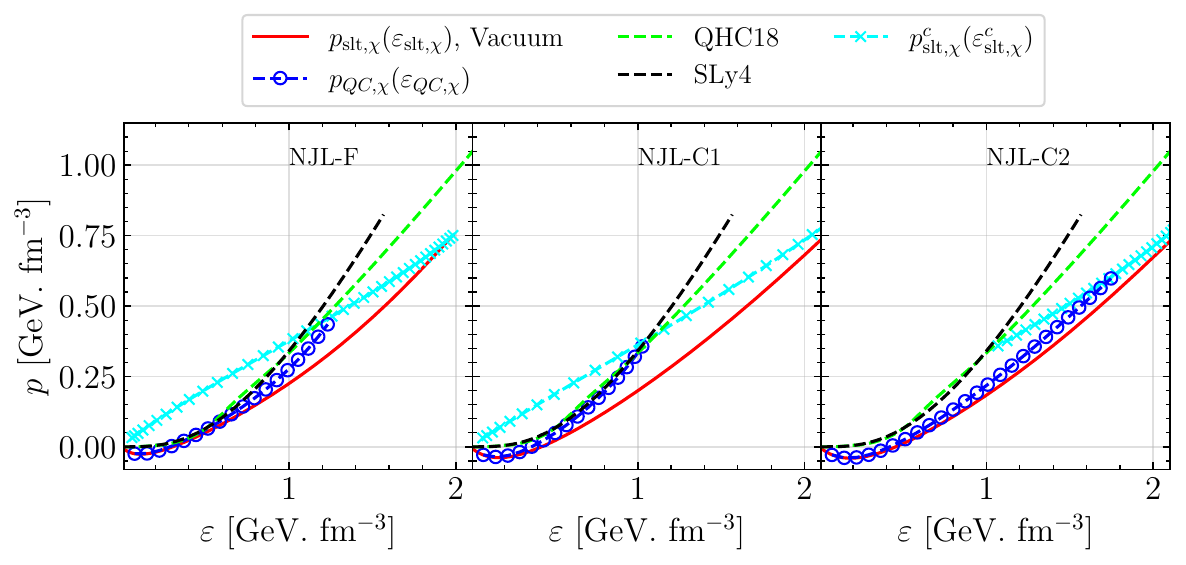}
    \caption{EoSs obtained under different formalisms are displayed. For reference, the evolution of the central pressure ($p_{\rm slt,\chi}^c$) as a function of the central energy density ($\varepsilon_{\rm slt,\chi}^c$) of the soliton is also shown. The NS EoS models are also displayed in dashed lines.}
    \label{fig:EoS_Eqc_compare}
\end{figure*}

To ensure consistency when comparing the resulting EoS Eq.~\eqref{eq:EoS_Eqc} with the NS EoS models as shown in Fig.~\ref{fig:EoS_Eqc_compare}, we rescale the energy density and pressure of Eq.~\eqref{eq:EoS_Eqc} using a vacuum renormalization factor $\chi$ to define,  
\begin{eqnarray}
\varepsilon_{QC,\chi}&\equiv& \chi\,\varepsilon_{QC}=M^*_N(s)\rho_N, \\
p_{QC,\chi} (\varepsilon_{QC,\chi}) &\equiv& {p_{\rm slt,\chi}(\varepsilon_{QC,\chi})} \, .
\end{eqnarray}
This rescaling implies that solitons with the renormalized energy density represent nucleons, such that in nuclear matter the quark cores (now replicating the nucleon) adjust to match the effective nucleon energy density ${\varepsilon}_{QC,\chi} \equiv M_N^* \rho_N$ with the in-medium effective nucleon mass $M^*_N$.

\begin{figure*}
    \centering
    \includegraphics[width=\linewidth]{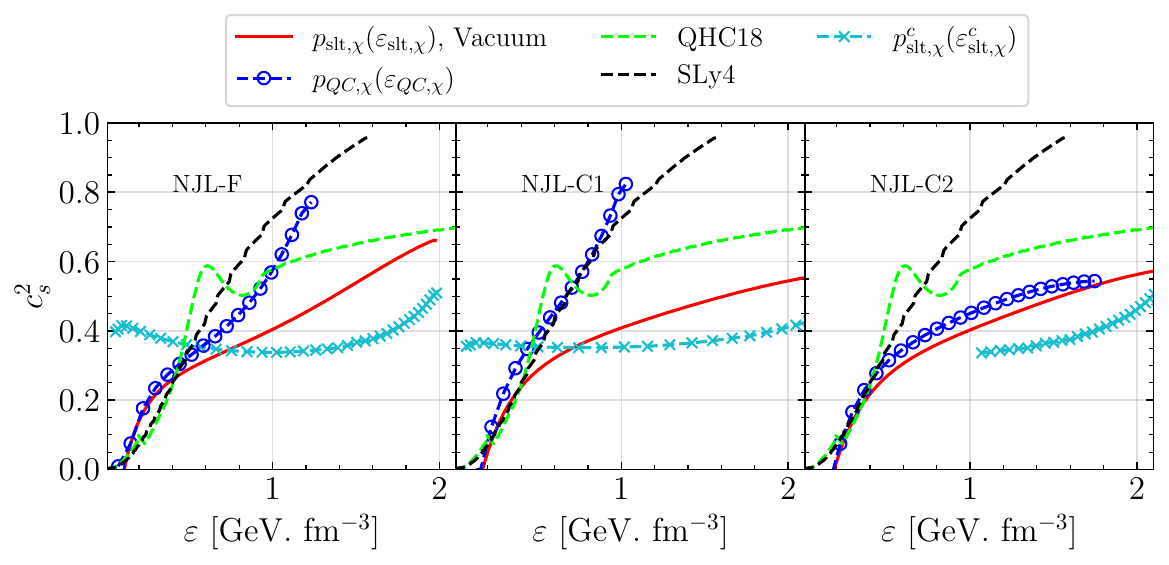}
    \caption{Squared speed of sound as a function of the energy density corresponding to the different EoS models shown in ~Fig.~\ref{fig:EoS_Eqc_compare}.}
    \label{fig:Cs2_energy_EQC}
\end{figure*}

With this procedure, a density-dependent EoS is constructed directly from microscopic soliton profiles. At low baryon densities $\rho_N$, the small bulk energy density $\varepsilon_{QC}$ corresponds to soliton configurations at large distances $r_{QC}$, where the local pressure becomes negative, as illustrated in Fig.~\ref{fig:EoS_Eqc_compare}. In Fig.~\ref{fig:Cs2_energy_EQC}, we display the evolution of the squared speed of sound ($c^2_s=\frac{dp}{d\varepsilon}$) for different EoSs of Fig.~\ref{fig:EoS_Eqc_compare}. The results of Fig.~\ref{fig:EoS_Eqc_compare} and Fig.~\ref{fig:Cs2_energy_EQC} demonstrate that a consistent treatment of chiral symmetry restoration produces a stiff soliton-based  EoS  (i.e, $p_{QC,\chi}(\varepsilon_{QC,\chi})$) compared to the vacuum soliton EoS (i.e, $p_{\rm slt,\chi}(\varepsilon_{\rm slt,\chi})$) and in quantitative agreement with established high-density NS models.

{When the interaction energy is taken into account, such that the quark degrees of freedom are readjusted to ensure that the  $\tilde{\varepsilon}_{QC,\chi}$ coincides with the nuclear matter energy density $\varepsilon_N$ at the nuclear density $\rho_N$, i.e.\ $\tilde{\varepsilon}_{QC,\chi}\equiv \varepsilon_N$ as given in ~Eq.~\eqref{eq:energy_density_nuclear}, a systematic shift in the EoS is observed. Specifically, this matching condition leads to a lower onset value of $\rho_N$ at which the soliton-based EoS ceases to be valid, thereby indicating an earlier transition to quark matter. This outcome stands in contrast to the approximation $\varepsilon_{QC,\chi}=M_N^*\rho_N$, where the interaction contribution is neglected, and which consequently predicts the termination of the soliton branch at higher densities. And given the energy density $\tilde\varepsilon_{QC,\chi}=\varepsilon_N$ at a density $\rho_N$, we define the corresponding pressure as $p_{\chi}(\varepsilon_N)=p_{\rm slt,\chi}(\varepsilon_N;s(\rho_N))$ and for instance this  is demonstrated in  Fig.~\ref{fig:EoS_Eqc_NJLF} for NJL-F parameter set. As one can notice with the re-definition of $\tilde{\varepsilon}_{QC,\chi}=\varepsilon_N$, the onset of the termination reduced to $6\rho_{\rm sat}$ from $9\rho_{\rm sat}$ resulting with the consideration of $ \varepsilon_{QC,\chi}=M_N^*\rho_N$.
}

For reference in Fig.~\ref{fig:EoS_Eqc_NJLF} we have also displayed the evolution of the soliton's central pressure ($p_{\rm slt}^c$) as a function of the central energy density  ($\varepsilon_{\rm slt}^c$). We remind that the central point is the least contaminated by surface effects. These central quantities also define a density-dependent EoS since the central nucleon density is a function of the scalar field $s$, which depends on the nuclear density. This is another prescription to obtain a density-dependent EoS.

\section{Conclusions}
\label{sec:Conclusions}

In this work, we began by investigating the vacuum properties of the soliton with baryon number $B=1$, as predicted by the NJL model through a hard-cutoff path-integral bosonization scheme. In this framework, we have made a detailed exploration of soliton properties for different NJL parameter sets. Our analysis shows that increasing the vector meson mass $M_v$ drives the $\omega$-field to become more localized in the soliton core, leading to a smaller iso-scalar charge radius. In contrast, the baryon charge radius and hard-core structure remain largely unaffected. This indicates that the valence quark distribution is robust against changes in the vector sector $M_v$, in contrast to the peripheral mesonic structure. We further found that the nucleon–scalar coupling is notably small, and the scalar susceptibility turns out to be negative. In addition, we compared the resulting core nucleon EoSs at vacuum across parameter sets, both with and without the scaling procedure used to reproduce the physical nucleon mass. While the unscaled EoSs closely reflect the underlying soliton dynamics, the application of the scaling factor $\chi$ systematically alters their stiffness, making them comparable to the high-density NS EoS models.

We then investigated the impact of chiral symmetry restoration on properties of solitons by varying the nuclear scalar field $s$ and solving for the corresponding soliton configurations in a fully self-consistent manner. Our results show that as the scalar field increases, quarks become progressively delocalized, leading to a broader baryon density profile and an enlarged hard-core radius of the nucleon. This indicates that nucleon cores begin to overlap at lower densities than expected from their vacuum radii, a feature with essential implications for the onset of repulsive effects in dense matter. The associated energy density further reveals a gradual decrease in the central density, consistent with reduced quark confinement. 

Furthermore, we demonstrated explicitly how chiral symmetry restoration modifies the interior dynamics of the soliton beyond the simple rescaling prescriptions adopted in earlier works. Once normalized to the vacuum nucleon mass, the in-medium EoS stiffens at scalar field values $s \sim 0.2f_\pi$–$0.3f_\pi$, reaching a level comparable to standard NS EoS models such as SLy4 and QHC18. Without this normalization, the desired stiffness is reached at larger scalar fields, $s \sim 0.5f_\pi$. The findings highlight how the nuclear medium scalar field drives both the swelling of the soliton core and the evolution of its EoS, providing a microscopic mechanism for medium modifications tied directly to chiral symmetry restoration. Our analysis shows that a self-consistent treatment 
captures the interplay between scalar and vector sectors and produce an EoS comparable to existing NS EoSs.

Within the mean-field treatment of symmetric nuclear matter in section~\ref{sec:soliton_and_density}, the effect of the surrounding medium on the soliton is encoded through uniform background fields, allowing the $\omega$ meson to asymptotically approach its in-medium mean-field value via a modified boundary condition. Implementing this prescription for the $\omega$ field in dense matter, in the same spirit as discussed in ~\cite{Li2024}, we find that there is no significant
change in the core properties of the soliton such as the soliton based EoS, hard-core radius. We find that with this modification, neither the qualitative behavior nor the quantitative results of our analysis are altered (see ~\ref{app:BC_dense}). For completeness, we have also examined an alternative implementation inspired by the Wigner–Seitz (WS) construction~\cite{Nagai2006}, in which the soliton is embedded in a finite spherical cell whose radius is fixed by the baryon density and the boundary conditions are imposed at the cell boundary. As discussed in ~\ref{app:BC_dense}, this scenario does not lead to stable soliton solution  within the present framework and therefore requires a more detailed and systematic investigation, which we defer to future work.

We have further investigated the properties of NJL-solitons embedded in nuclear matter through a self-consistent framework that accounts for the effects of chiral symmetry restoration at finite density. By describing nuclear matter within the relativistic mean-field (RMF) formalism, where the mesonic interactions are derived from a bosonization approach consistent with the adopted soliton model from the underlying NJL model, we have explored the density-dependent modifications to soliton characteristics, such as charge radii, as well as their implications for the EoS. This approach allows us to explore the density-dependent EoS, revealing a gradual modification of nucleon structure in dense matter. The observed overlap of soliton cores and the corresponding breakdown of the soliton-based EoS at high densities point to the onset of hard-deconfinement at density $\sim 8\rho_{\rm sat}-10 \rho_{\rm sat}$ depending upon the parameters, providing a physically transparent connection between in-medium chiral dynamics and the possible emergence of quark matter. The results presented in section~\ref{sec:soliton_and_density} are the artifact of the slow evolution of the scalar field with the density due to the low scalar couplings. We have also presented an alternative way to extract an EoS from the different energy-density and pressure profiles inside the nucleon by considering the central point $r=0$ obtained for different values of the scalar field $s$ fixed by the nuclear density.

The soliton properties presented in this work can be further improved by means of several systematic extensions. In particular, one may consider the contributions of the axial meson in the NJL-soliton description, the effects of spin-isospin quantization, the role of the scalar field (whose slow spatial variation across the soliton has been neglected here), and the inclusion of an explicit confining potential. Extending the analysis beyond SNM to $\beta$-equilibrated neutron-star matter or pure neutron matter, as well as moving beyond the mean-field approximation toward a complete Hartree-Fock treatment, would also provide a more comprehensive understanding of the role of chiral symmetry restoration in the results presented in section~\ref{sec:soliton_and_density}. Furthermore, alternative treatments of nuclear matter with explicit quark degrees of freedom and finite-size corrections~\cite{Bentz2025,JAMINON1989}, such as simulating solitons in a Wigner-Seitz cell or employing a Skyrmion lattice crystal approach~\cite{Ma2013,Ma2014,Dong2013}, may yield a different evolution of the scalar field and consequently lead to further refinements of the soliton properties and the resulting EoS.

\begin{figure}[t]
\centering
\includegraphics[width=\linewidth]{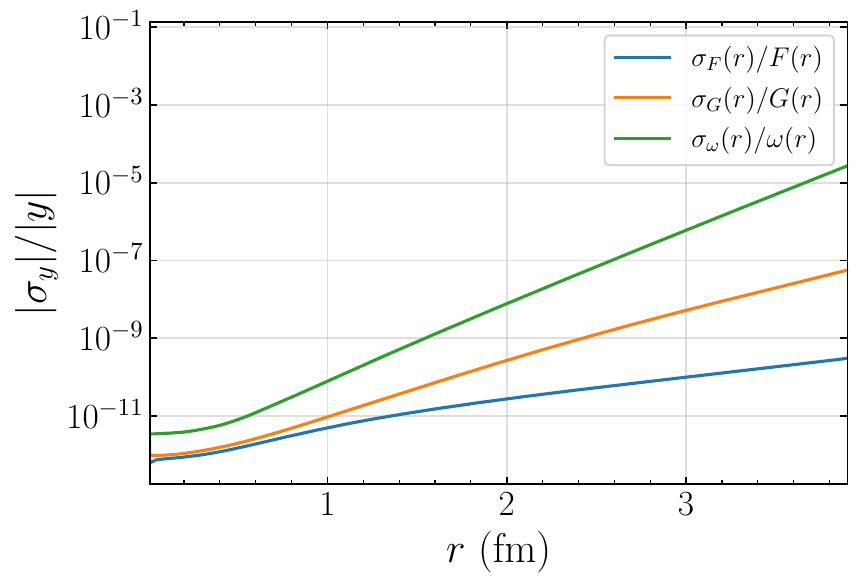}
\caption{$\sigma_y/y$ obtained using ~Eq.~\eqref{eq:sigma_i}. }
\label{fig:sigma_hybrid}
\end{figure}

\section{Acknowledgment}
The authors acknowledge the support of the CNRS-IN2P3 MAC masterproject, the project RELANSE ANR-23-CE31-0027-01 of the French National Research Agency (ANR), the European Union’s Horizon 2020 research and innovation program under grant
agreement STRONG–2020-No824093.
This work makes use of \texttt{NumPy} \cite{vanderWalt:2011bqk}, \texttt{SciPy} \cite{Virtanen:2019joe}, \texttt{Matplotlib} \cite{Hunter:2007}, \texttt{jupyter} \cite{jupyter} software packages, the Gnu Scientific Library software and reference manual \cite{galassi2018scientific}.

\section{Data and Code Availability Statement}
The  computational codes used to 
support the findings of this study are available upon reasonable request.  The datasets generated during and/or analysed in  the current study are publicly  available at \href{https://github.com/bikramp-hub/NJL-Chiral-Soliton/}{GitHub/NJL-Chiral-Soliton}.] 
\appendix
\section{Error Analysis}\label{app:error}

\begin{figure*}[t]
\centering
\begin{subfigure}{.49\textwidth}
  \centering
  \includegraphics[width=\linewidth]{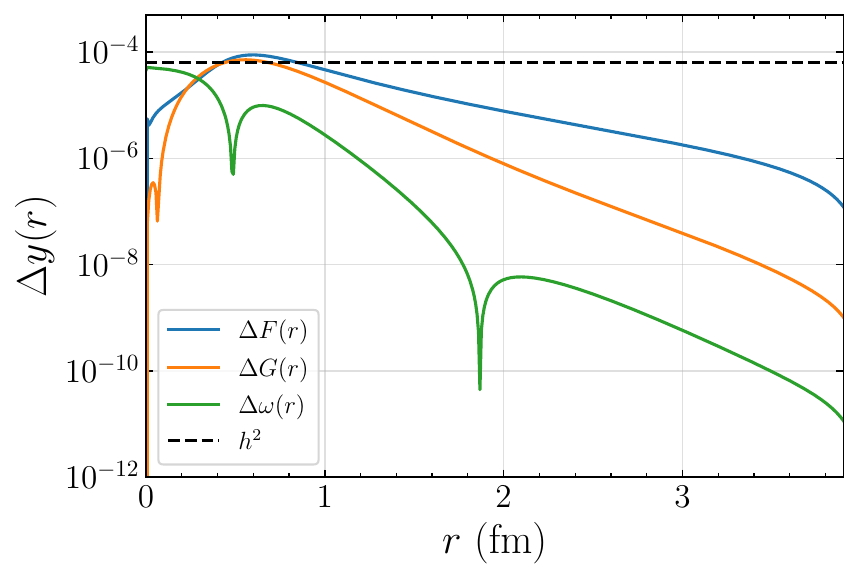}
   \caption{} 
   \label{fig:error_inF}
\end{subfigure}%
\begin{subfigure}{.49\textwidth}
  \centering
  \includegraphics[width=\linewidth]{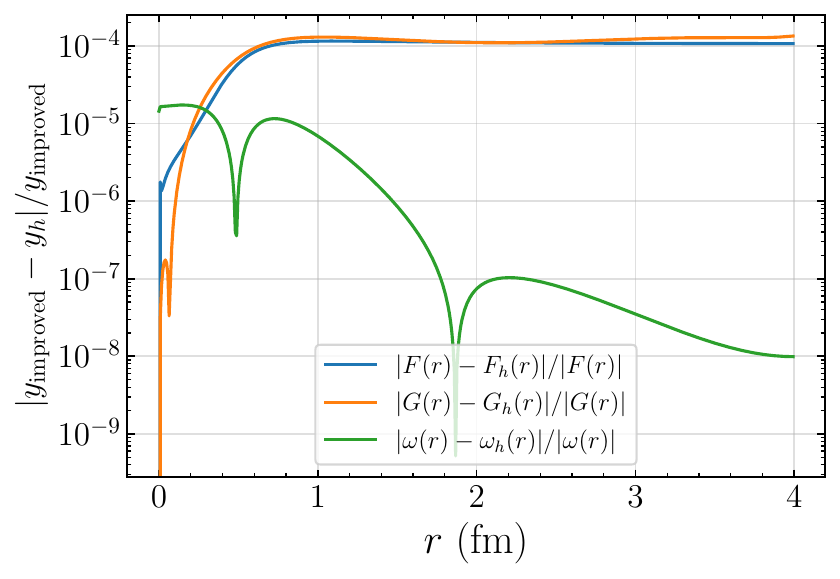}
  \caption{}
  \label{fig:relative_error}  
\end{subfigure}
\caption{(a) Absolute errors in the field profiles obtained via the relaxation method. (b) Corresponding relative errors compared to the improved Richardson-extrapolated solutions.}
\end{figure*}

In our numerical implementation of the relaxation method, the possible sources of error arise from two main factors: (i) discretizations due to finite differences, and (ii)  by the root-finding algorithm that minimizes the residuals (i.e., the difference between the left-hand side and right-hand side) of the field equations in Eqs.~\eqref{eq:d2F_NJL}-\eqref{eq:d2w_NJL}. The errors on the numerical solutions $y = [F(r), G(r), \omega(r)]$ due to the minimization procedure can be estimated using the upper bound error formula from Ref.~\cite{Keller1974}:
\begin{equation}\label{eq:sigma_i}
\sigma_i = \epsilon^{1/2} \frac{||\text{Res}(y)||}{||J(y) \cdot e_i||},
\end{equation}
where $\epsilon \sim \text{Ftol} \sim \text{tol}$ is the stopping criterion tolerance, $||\cdot||$ denotes the 2-norm, $J$ is the Jacobian matrix, and $e_i$ is the $i^\text{th}$ column of the identity matrix. The evolution of $\sigma_i$ for $N = 500$, $\epsilon = \text{tol} = 10^{-9}$, and boundary at $r_\infty = 4$ fm using the parameter set of Ref.~\cite{Fukushima2020} is shown in Fig.~\ref{fig:sigma_hybrid}.

The discretizations error arises from evaluating the residuals of the equations using central finite differences, which introduce a local truncation error of $\sim \mathcal{O}(h^2)$ in the derivatives $F^{\prime}(r), G^{\prime}(r)$, and $\omega^{\prime}(r)$. For linear problems, the global error remains $\sim \mathcal{O}(h^2)$~\cite{kelley1995iterative,Richard2010}. However, for nonlinear coupled systems, this error behavior is not always guaranteed.

To estimate the discretization error and to compare with a higher-accuracy solution, we employ Richardson extrapolation~\cite{Richard2010}. The error in the numerical solution $y_h(r)$ with step size $h$ is approximated by
\begin{equation}
\Delta y(r) = y_h(r) - y(r) \approx \frac{4}{3} \left( y_{h/2}(r) - y_h(r) \right),
\end{equation}
while an improved solution accurate to $\mathcal{O}(h^4)$ is given by
\begin{equation}
y_{\text{improved}}(r) = \frac{4 y_{h/2}(r) - y_h(r)}{3}.
\end{equation}

The absolute errors in the field profiles and their relative errors with respect to the improved solutions, computed within the extrapolation approach for the parameters of Ref.~\cite{Fukushima2020}, are shown in Fig.~\ref{fig:error_inF} and \ref{fig:relative_error}, respectively. For comparison, the $h^2$ behavior is also displayed in~Fig.~\ref{fig:error_inF}. Using $N = 500$, $r_\infty = 4$ fm, and $h \sim 10^{-2}$ fm, the numerical errors in the model observables are found to be:
\begin{align*}
    \Delta M_N^{\rm model} &\sim 10^{-4}~\text{MeV}, \\
    \Delta \sqrt{\langle r_B^2 \rangle} &\sim 10^{-5}~\text{fm}, \\
    \Delta \sqrt{\langle r_S^2 \rangle} &\sim 10^{-4}~\text{fm}.
\end{align*}

\begin{figure}[t]
    \centering
    \includegraphics[width=\linewidth]{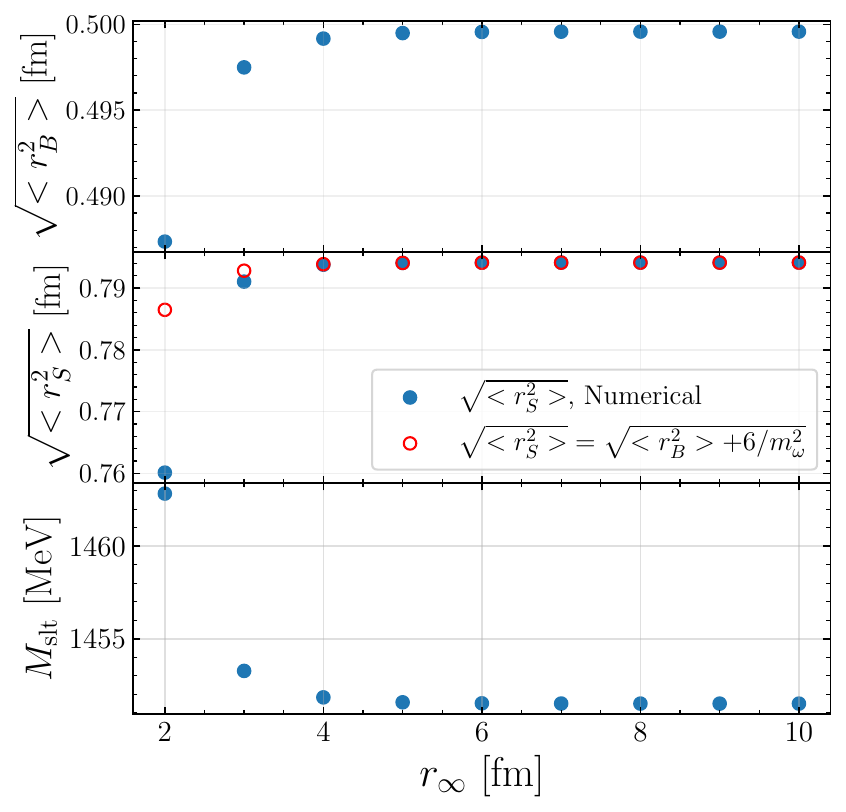}
    \caption{Dependence of soliton properties on $r_\infty$ for the relaxation method. Upper Panel: Variation of $\sqrt{\langle r_B^2 \rangle}$. Middle panel: $\sqrt{\langle r_S^2 \rangle}$. Lower Panel: Variation of soliton mass $M_{\rm slt}$ for different choices of $r_\infty$. All results correspond to the soliton properties obtained with meson parameters of Ref.~\cite{Fukushima2020}.}
    \label{fig:rb_rs_Mn_vsrinf_relax}
\end{figure}

To improve convergence and properly handle the boundary conditions at $r \to \infty$, we initially solve the equations in the compactified coordinate $t = \tanh(r)$. However, for precise physical observables, we re-solve the equations in the physical coordinate $r$ using the relaxation method, starting from the interpolated solution in $t$-space. The results presented in the main text correspond to a fine grid spacing $h \sim 10^{-3}$ fm, which provides improved precision over the $h \sim 10^{-2}$ fm case discussed here. Furthermore, we find that soliton observables are relatively insensitive to the choice of $r_\infty$ beyond $r_\infty > 4$ fm, as shown in Fig.~\ref{fig:rb_rs_Mn_vsrinf_relax}.

\section{Nucleon-Scalar coupling $g_s$ and vector properties of the NJL model}
\label{app:gs_and_mv}

\begin{figure}[t]
    \centering
    \includegraphics[width=\linewidth]{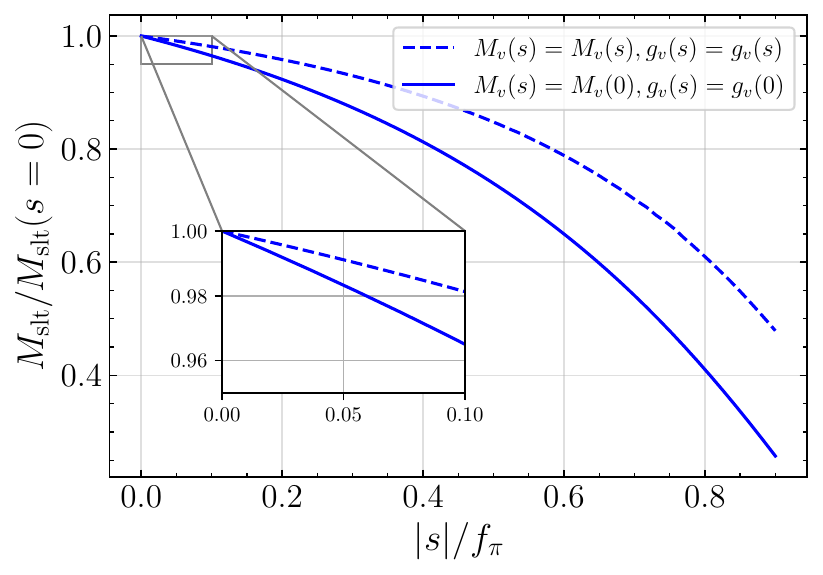}
    \caption{Evolution of the soliton (nucleon) mass $M_N$ with the nuclear scalar field $s$ for the NJL-F parameter set. Solid (dashed) line: vector properties fixed at vacuum (evolved with $s$).}
    \label{fig:Ms_vs_s}
\end{figure}

Following the definition of Eq.~\eqref{eq:ener}, the soliton mass $M_{s}(s)$ can be evolved by allowing both scalar-sector parameters ($F_{\pi}, M_{\pi}$) and vector-sector parameters ($g_v, M_v$) to vary with $s$. In this self-consistent treatment, the nuclear scalar coupling $g_s$ receives positive contributions from the scalar sector and negative contributions from the vector sector, leading to a reduced overall value.  

In contrast, several works, see Refs.~\cite{CHRISTOV1990,Alkofer1991}, incorporate in-medium chiral symmetry restoration only through the scalar properties ($M_{\pi}, F_{\pi}, M_{\sigma}$), effectively treating the scalar field as the sole mediator of restoration while keeping the vector sector fixed at its vacuum values. For clarity, we distinguish two cases:
\begin{enumerate}
    \item \textbf{Case 1:} mass of the soliton $M_{\rm slt}(s)$  evaluated following section~\ref{sec:Method_NJL}, with $M_v$ and $g_v$ fixed at their vacuum values (solid line in~Fig.~\ref{fig:Ms_vs_s}).
    \item \textbf{Case 2:} mass of the soliton $M_{\rm slt}(s)$  evaluated following section~\ref{sec:Method_NJL} with both $M_v(s)$ and $g_v(s)$ evolving with $s$ (dashed line in~Fig.~\ref{fig:Ms_vs_s}).
\end{enumerate}

As shown in~Fig.~\ref{fig:Ms_vs_s}, neglecting vector-sector modifications (Case~1) yields $g_s = 5.02$, $5.47$, and $4.33$ for the NJL-F, NJL-C1, and NJL-C2 parametrizations, respectively. In the main text, however, we adopt the self-consistent Case~2, where both scalar and vector parameters evolve with $s$, to describe the in-medium modifications of soliton properties.

\section{In-Medium Meson Properties}
\label{app:mesons_in_nuclear_matter}
\begin{figure*}[htbp]
\centering
\includegraphics[width=\linewidth]{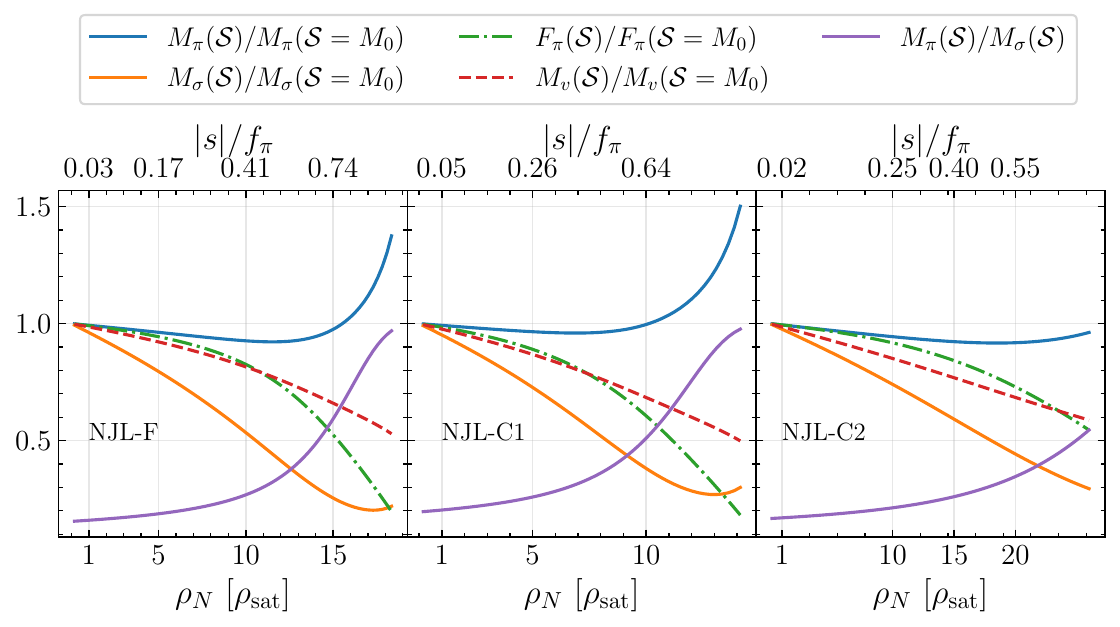}
\caption{Relative changes of in-medium meson properties, normalized to their vacuum values, as functions of nuclear matter density ($\rho_N$) and the corresponding nuclear scalar field $s$. We have also shown  evolution of the ratio among the pion and the scalar meson masses with  density.}
\label{fig:meson_props_in_matter}
\end{figure*}

We display in-medium meson properties in nuclear matter for different NJL parameterizations in~Fig.~\ref{fig:meson_props_in_matter}. For a given baryon density $\rho_N$, we first determine the in-medium nuclear scalar field $s$ (equivalently ${\cal S}(s)$) using~Eq.~\eqref{eq:nuclear_s_field}, as described in section~\ref{sec:soliton_and_density}. From this, the corresponding in-medium meson properties   $M_{\pi}({\cal S})$, $F_{\pi}({\cal S})$, $M_{\sigma}({\cal S})$, and $M_v({\cal S})$ are obtained via Eqs.~\eqref{eq:Mpi}, \eqref{eq:Fpi}, ~\eqref{eq:Msigma}, and \eqref{eq:Mv}, respectively,  following the method outlined in section~\ref{sec:Method_NJL}. For clarity, all quantities are normalized to their vacuum values, so that the plots highlight their relative changes with density. As the density increases and chiral symmetry is progressively restored ($|s|\to f_{\pi}$), the in-medium pion mass $M_{\pi}({\cal S})$ approaches the in-medium scalar mass $M_{\sigma}({\cal S})$. 

\section{Effect of Boundary Conditions for the Soliton in Dense Matter}
\label{app:BC_dense}

As discussed in section~\ref{sec:soliton_and_density}, the density evolution of the soliton is driven by the scalar mean field $\cal{S}$ generated by the surrounding medium, which is treated as uniform in the mean-field approximation. In symmetric nuclear matter, the pion and $\rho$-meson mean fields vanish outside the soliton, so their boundary conditions remain identical to the vacuum case. The $\omega$ meson, however, asymptotically can acquire a nonzero in-medium expectation value (see Ref.~\cite{Li2024}).

In uniform matter with baryon density $\rho_N$, the $\omega$ mean field satisfies
\begin{equation}
\bar{\omega} = \frac{g_v}{M_v^2}\,\rho_N ,
\end{equation}
and the soliton profile may asymptotically approach
\begin{equation}
\omega_{\rm soliton}(r)\big|_{r\to r_\infty} = -\bar{\omega} .
\end{equation}
This prescription corresponds to treating the soliton as a localized excitation embedded in infinite dense matter, with boundary conditions imposed at $r_\infty \to \infty$, as discussed for instance in Ref.~\cite{Li2024}. We refer to this choice as \textbf{B.C-2}, while the vacuum-like boundary conditions used throughout the main text are denoted as \textbf{B.C-1}. Explicitly,
\begin{align}
\text{B.C-1:}\quad &
\omega'(0)=0,\quad \omega(r_\infty\to\infty)=0 , \\
\text{B.C-2:}\quad &
\omega'(0)=0,\quad \omega(r_\infty\to\infty)=-\bar{\omega}(\rho_N) ,
\end{align}
with identical conditions for the $F$ and $G$ profiles in both cases.

\begin{figure*}[htbp]
\centering
\begin{subfigure}{0.49\textwidth}
\includegraphics[width=\linewidth]{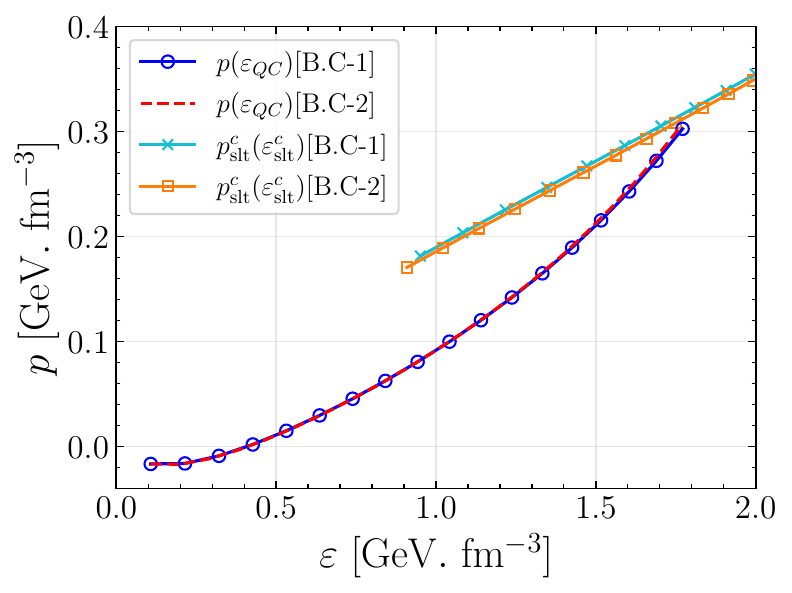}
\caption{ }
\label{fig:EoS_BC}
\end{subfigure}
\begin{subfigure}{0.49\textwidth}
\includegraphics[width=\linewidth]{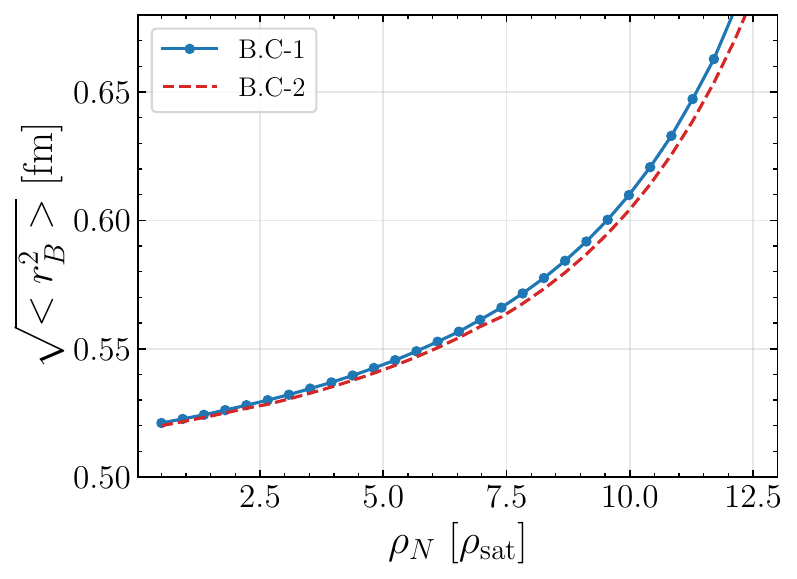}
\caption{}
\label{fig:RB_BC}
\end{subfigure}    
\caption{(a) EoSs obtained under different B.Cs are displayed. For reference, the evolution of the central pressure ($p_{\rm slt}^c$) as a function of the central energy density ($\varepsilon_{\rm slt}^c$) of the soliton is also shown. (b) Evolution of the hard-core radius as a function of density obtained under different B.Cs. The comparison is performed for the NJL-F parameterization.}
\label{fig:compare_EoS_BC}
\end{figure*}

We have verified that implementing \textbf{B.C-2} does not lead to any significant modification of the soliton core properties. In particular, the soliton-based equation of state ~Eq.~\eqref{eq:EoS_Eqc}, the EoS extracted from the soliton center, the density evolution of the hard-core radius, and the truncation density of soliton based EoS remain essentially unchanged, as shown in Fig.~\ref{fig:compare_EoS_BC}. 

\begin{figure}[htbp]
    \centering
    \includegraphics[width=\linewidth]{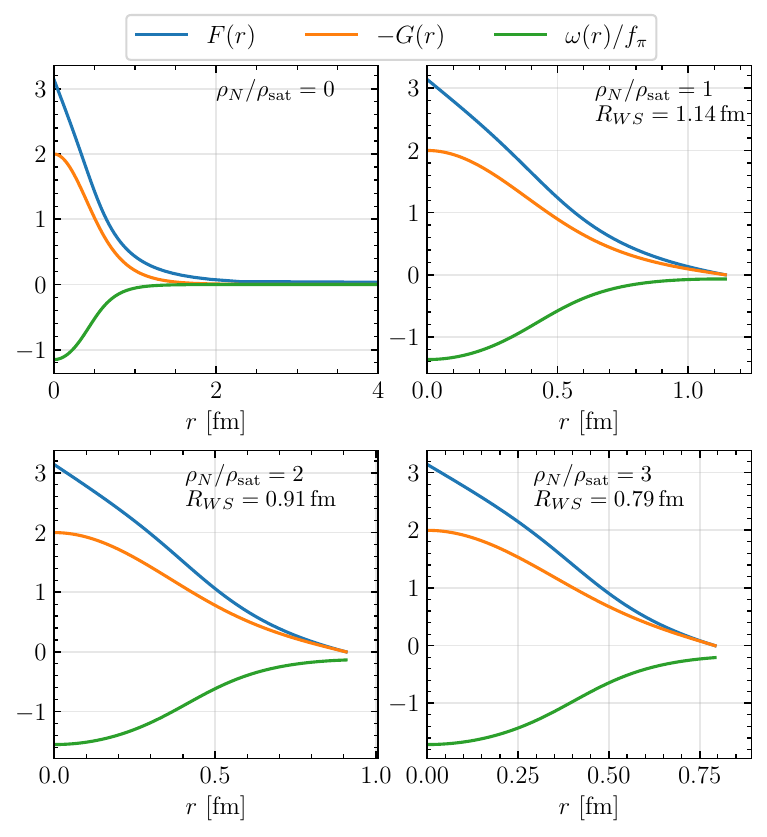}
    \caption{Meson profiles of the soliton at different bulk baryon densities under B.C-3. The comparison is performed for the NJL-F parameterization.}
    \label{fig:meson_profiles_B.C3}
\end{figure}

\begin{figure}[htbp]
    \centering
    \includegraphics[width=\linewidth]{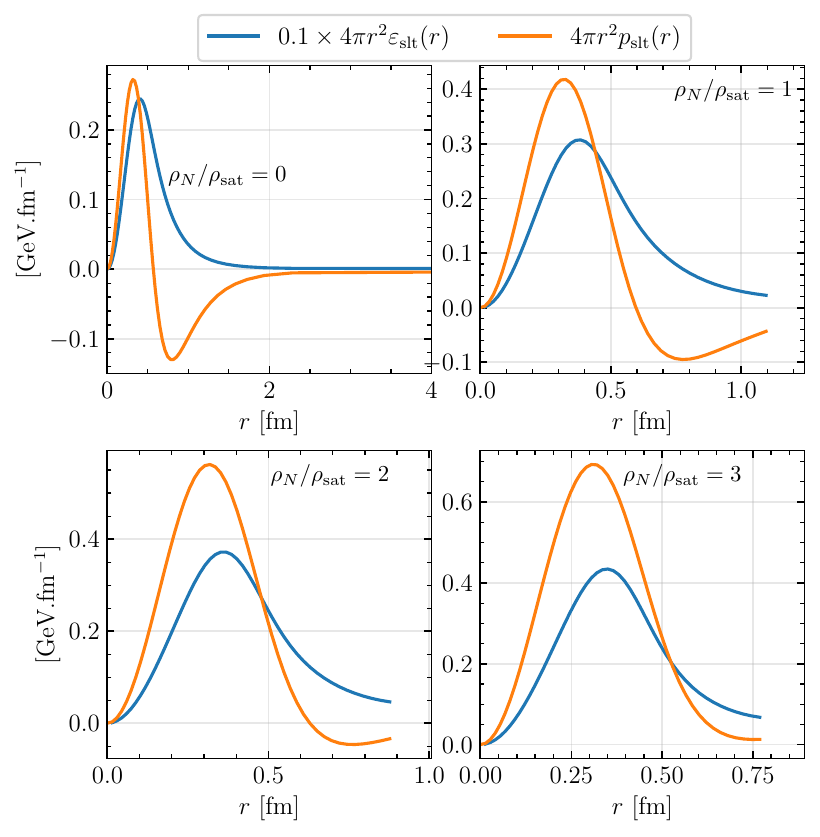}
    \caption{ Distribution of the soliton's  energy density and pressure at different bulk nuclear matter densities. The comparison is performed for the NJL-F parameterization.}
    \label{fig:density_profiles_B.C3}
\end{figure}

For completeness and to comment, we also explored an alternative construction inspired by Wigner-Seitz treatments of dense matter~\cite{Nagai2006}, where the soliton is placed at the center of a spherical unit cell of radius
\begin{equation}
R_{\rm WS} = \left(\frac{4\pi}{3}\rho_N\right)^{-1/3} ,
\end{equation}
and boundary conditions are imposed at $r_\infty=R_{\rm WS}$. We denote this choice as \textbf{B.C-3}, with
\begin{equation}
\omega'(0)=0,\quad \omega(R_{\rm WS})=-\bar{\omega}(\rho_N) .
\end{equation}
 Representative meson profiles and energy-pressure distributions obtained under \textbf{B.C-3} are shown in Fig.~\ref{fig:meson_profiles_B.C3} and Fig.~\ref{fig:density_profiles_B.C3} respectively. In this case, we find that the stability condition $\int d^3r\, p(r)=0$ is violated once $R_{\rm WS}\lesssim 1~\mathrm{fm}$, indicating that this prescription is not compatible with a stable soliton solution in the present NJL-based framework. A thorough investigation of such modified boundary conditions would require a separate study, similar to analyses performed for Skyrmion crystals or solitons on a lattice.

 We therefore conclude that the boundary conditions adopted in the main text (\textbf{B.C-1}) are the most consistent choice within the mean-field and Born-Oppenheimer approximations employed in this work. Furthermore, a similar treatment for the scalar field profile could be implemented in future work by going beyond the Born-Oppenheimer approximation. In such an extended framework, the soliton would receive additional medium-induced contributions from both scalar and vector mesons, together with consistently modified boundary conditions, allowing for a more complete description of in-medium soliton dynamics.

\bibliographystyle{spphys}       
\bibliography{draft}

\end{document}